\pdfoutput=1

\documentclass[12pt,a4paper]{report}

\usepackage{cite}
\usepackage{fancyvrb} 
\usepackage{graphicx} 
\usepackage{makeidx}
\usepackage{url} 
\usepackage{listings}
\usepackage{amsmath}
\usepackage{subfiles}
\usepackage{paralist}
\usepackage{paralist}
\usepackage{listings}
\usepackage{setspace}
\usepackage{pdfpages}
\usepackage[T1]{fontenc}
\usepackage[utf8x]{inputenc}
\renewcommand{\and}{\hspace{.5cm}} 
\newcommand{\bemph}[1]{\textbf{\emph{#1}}}

\fvset{baselinestretch=1}

\lstdefinestyle{mystyle}{
    basicstyle=\footnotesize\singlespacing,
    breakatwhitespace=false,
    breaklines=true,
    captionpos=b,
    keepspaces=true,
    numbers=left,
    numbersep=5pt,
    showspaces=false,
    showstringspaces=false,
    showtabs=false,
    tabsize=2
}

\newcommand{\researchTitle}{Towards Curating Social Media Data}
\urldef{\mails}\path|{z5077732}@cse.unsw.edu.au|

\begin{document}

\title{ \researchTitle } \author{Kushal Vaghani \\ \\
School of Computer Science \& Engineering \\ University of New South
Wales \\ Sydney, Australia.}

\includepdf[pages={1}]{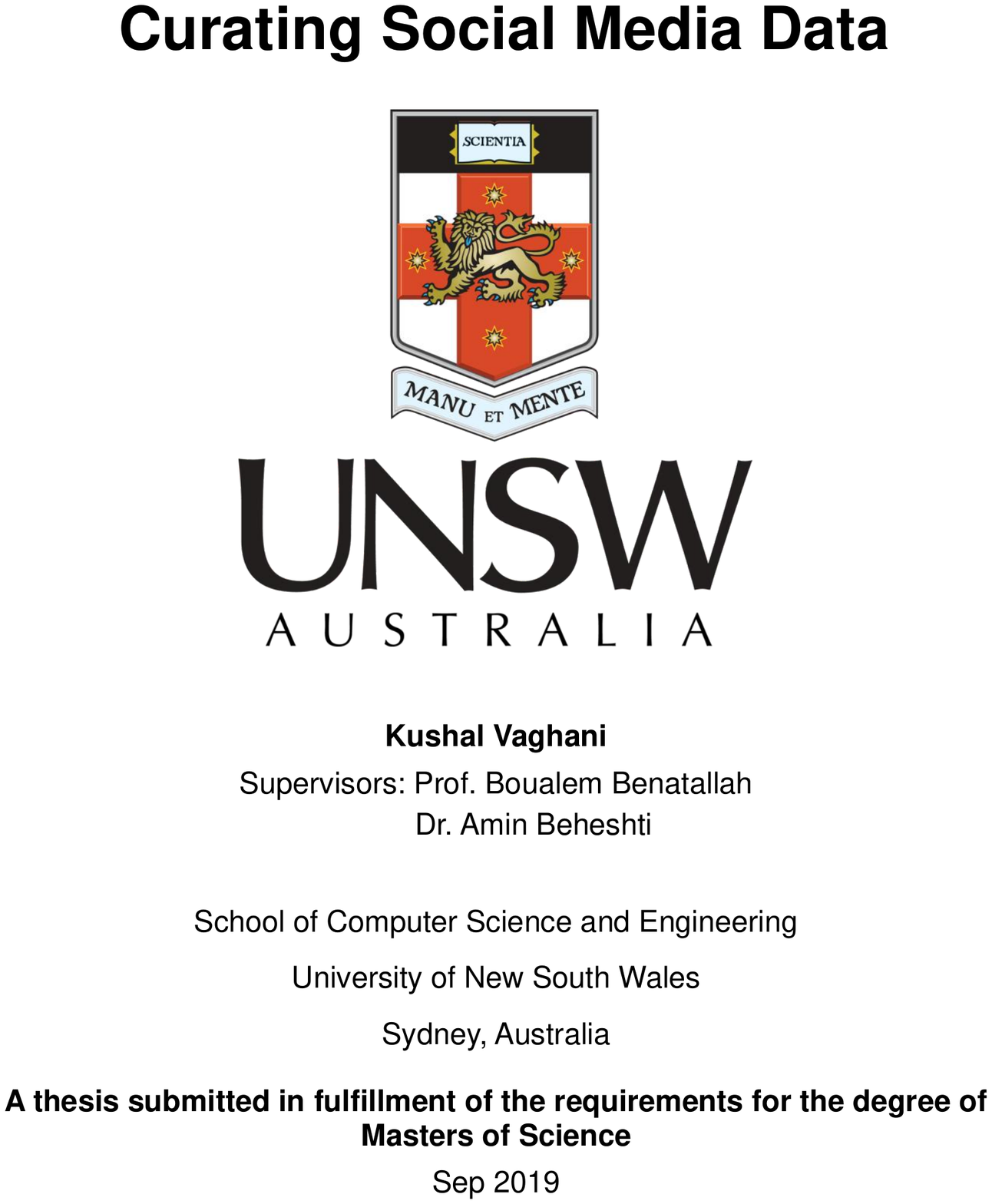}

\chapter*{Acknowledgements}
\addcontentsline{toc}{chapter}{Acknowledgements}
First of all, I would like to thank my supervisors Prof. Boualem Benatallah and Dr. Amin Beheshti for
their guidance and encouragement during my research. Without their constant support this work
would not have been possible. Moreover, I would also like to thank them, for allowing me to work with the outstanding team of researchers at UNSW. I sincerely believe, I have grown both as an individual and researcher under their supervision.

I am thankful to everyone in the Service-Oriented Computing (SOC) group at UNSW, especially Mr. Alireza Tabebordbar for his friendship, support and helping me in all possible ways to evaluate my research work.
In addition, I would like to thank the review panels and the anonymous reviewers who provided suggestions and helpful feedback on my publications and thesis. Also, I hereby acknowledge Dr. Amin Beheshti, Prof. Boualem Benatallah and Alireza Tabebordbar for their valuable contribution as co-authors for the research paper publication.

In addition, I would like to thank administrative and technical staff members of the school of computer science and engineering at UNSW who have been kind enough to advise and help in their respective roles.

My final thanks go to my family for supporting me in pursuing an advanced
degree. I am especially grateful to my daughter Ziva for her encouragement during my
study in UNSW. I faced a terrible loss due to my father's demise during the tenure of this research. I hope he would be happy seeing me completing my research.
\\\\
Kushal Vaghani
\\
Sydney, Australia
\\
July, 2019

\chapter*{Abstract}
\addcontentsline{toc}{chapter}{Abstract}

 Social media platforms have empowered the democratization of the pulse of people in the modern era. Due to its immense popularity and high usage, data published on social media sites (e.g., Twitter, Facebook and Tumblr) is a rich ocean of information. Therefore data-driven analytics of social imprints has become a vital asset for organisations and governments to further improve their products and services. However, due to the dynamic and noisy nature of social media data, performing accurate analysis on raw data is a challenging task. A key requirement is to curate the raw data before fed into analytics pipelines. This curation process transforms the raw data into contextualized data and knowledge.
 
 We propose a data curation pipeline, namely   CrowdCorrect, to enable analysts cleansing and curating social data and preparing it for reliable analytics. Our pipeline provides an automatic feature extraction from a corpus of social media data using existing in-house tools. Further, we offer a dual-correction mechanism using both automated and crowd-sourced approaches. The implementation of this pipeline also includes a set of tools for automatically creating micro-tasks to facilitate the contribution of crowd users in curating the raw data. For the purposes of this research, we use Twitter as our motivational social media data platform due to its popularity.

\tableofcontents

\chapter{Introduction}

This chapter is organized as follows. In
Section~\ref{sec:intro_background}, we introduce the basic background on social media data curation. In
Section~\ref{sec:intro_motivation}, we outline the research problem that we are
addressing and discuss the motivations. In Section~\ref{sec:intro_contributions},
we summarize our contributions. Section~\ref{sec:intro_thesis_organization}
provides the organization of this thesis.

\section{Background}
\label{sec:intro_background}

Ever since the dawn of the industrial age, understanding of data to gain knowledge and wisdom has been given utmost importance. \emph{Data} can be termed as merely a collection of facts such as numbers, words, measurements and posts on blogs~\cite{bellinger2004data}. Today, the continuous improvement in connectivity, storage and processing allows access to data deluge from open and private data sources. Such raw data needs to be processed to increase its usefulness. Once raw data is processed and transformed into knowledge; this can help achieve meaningful insights and decision-making processes.

With the modern popularity of social media networks such as Twitter\footnote{https://twitter.com/}, Facebook\footnote{https://www.facebook.com/} and LinkedIn\footnote{https://au.linkedin.com/}; an enormous amount of open data content (e.g., tweets on Twitter) is published on a daily basis~\cite{stieglitz_social_2018}. As an example, there are approximately 500 million tweets posted each day on Twitter\footnote{https://www.omnicoreagency.com/twitter-statistics/}. It is no secret~\cite{SocialDataMap} that the world is glued to social media with user populations in millions. The data within these social channels natively captures the pulse and opinions of the masses in a way never before available. This opens up new opportunities for deeper understanding of several aspects such as trends, opinions and influential actors. Such data can provide valuable insights to aid decision making in diverse areas such as marketing, public policy and healthcare. Organisations can use social data to target and validate their marketing campaigns; governments can device better policies and improve their services. As an example, the Australian government's Department of Jobs and Small Businesses Website\footnote{https://www.jobs.gov.au/social-media-usage-and-policies} articulates that it uses social media to improve stakeholder engagement amongst other items such as countering inaccurate news and promoting transparency. Another research 
study~\cite{erdougmucs2012impact}, links social media usage to brand and product loyalty. Organisations as well as governments, therefore, consider analysis of such information as a vital asset and a strategic priority.

Raw data from social media sites needs to be pre-processed, contextualized and prepared (i.e., curated) for analytics. Motivations for curating social media data are discussed in Section 1.2. The curation process consists of ingesting, cleaning, merging, linking, enriching and preparing the data for analytics. In short, it transforms the raw (structured, semi-structured and unstructured) data into curated data, i.e., contextualized data and knowledge~\cite{CurationAPI}. This curated data is then made available to applications and end-users.

\section{Motivations and Problems}
\label{sec:intro_motivation}

\begin{figure}[h!]
\centering
\includegraphics[width=\textwidth,height=\textheight,keepaspectratio]{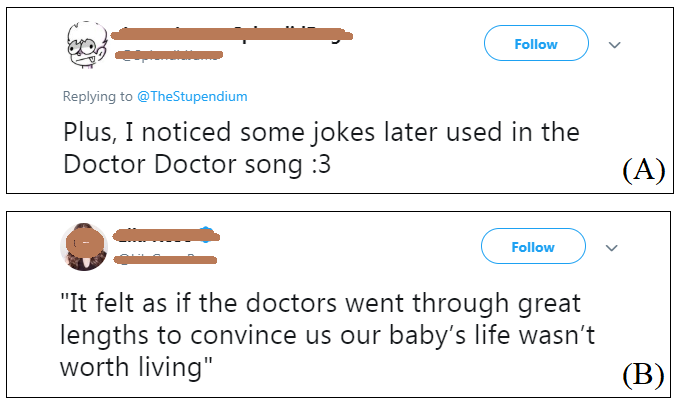}
\caption{Tweet Example: Same word ``doctor'' used in different context.}\label{fig:social-motivations}
\end{figure}

There are several motivations and research issues in preparing the raw social data for analytics tasks. Raw data from social platforms is generally semi-structured; consisting of unstructured parts such as text and media with some structured parts such as friend/follower relationships. Structured data is organised and easy to process, e.g., list of followers on Twitter in the standard JSON\footnote{https://www.json.org} format; while unstructured data is difficult to process~\cite{katal2013big}. Next, since the social networks allow their users to express themselves without any restrictions (e.g., freeform text in tweet text as an example); there is a high amount of noise in the raw data~\cite{soto_data_nodate,eisenstein2013bad}. Such noise includes misspellings, slang words, abbreviations, truncations, incorrect syntax, grammatical errors. Table~\ref{social-text} illustrates some examples of such noise prevalent in the social medium. In addition, texts or words expressed often, need proper contextualization to be comprehended and relevant. For example, Figure~\ref{fig:social-motivations} shows two tweets taken from Twitter which contain the word \emph{doctor}. Whilst the second tweet (B) is related to health; the first tweet (A) refers to a song.

\begin{table}
\centering
\begin{tabular}{ll}
 \textbf{Issue} & \textbf{Example}\\ \hline
 Spelling Mistakes & e.g., healht, hspital  \\
 Abbreviations & e.g., lol (laugh out loud), aust.(Australia)\\
 Phonetic subsitutions & e.g., lyk (like), 2 (to)      \\
 Jargons & e.g., pill (for medicine) \\
 Truncation & e.g., tom (for tomorrow)   \\
 Deletion of words & e.g., gng home (for 'I am going home')
\end{tabular}
\caption{Common text issues in social media}
\label{social-text}
\end{table}

Essentially the quality of the raw social data is low~\cite{immonen2015evaluating}; which introduce linguistic challenges in machine processing for analytics and can lead to inaccurate analysis~\cite{adedoyin-olowe_survey_nodate}. These quality issues are further compounded due to large volumes of data generated daily (size) at a continuous rate (i.e., dynamism). Without a robust data cleansing and curation process to resolve these issues; the results of carrying out analytics would be erratic. Here, cleansing refers to improving the quality of raw data; while curation refers to a process to produce contextualized data which can be used
for analytics~\cite{CurationAPI,palmer2013foundations}.

Research has shown that state-of-art natural language processing (NLP) systems perform significantly worse on social media 
text~\cite{eisenstein2013bad}.
In order to better understand these challenges, we consider a motivating scenario in Twitter.

\subsubsection{An Example from Twitter}
\label{sub:twitter_example}

Shown below are three tweets extracted from a corpus of tweet data in public response to Australian government's annual budget announcement in 2016. While there are thousands of similar tweets in understanding raw social data, the tweets illustrated below highlight some of the challenges with social media data:

\begin{compactenum}[a)]
	\item \emph{"MRI and CT Scan must be at loooowest \$ for needy patients \#budget"}
     \item \emph{"Healht insurers given all clear. OMG!! https://t.co/ew95fo9dn"}
	\item \emph{"@arcgp: low socio-economic bypass pat. head to emergcy departments as aussie govt's budget freezes \#budget2016"}
\end{compactenum}

A few issues are apparent from the above sample tweets. First, there is no preference for any standard form of text or language. That is, such social medium data is full of grammatical and syntactical errors. Second, there is a heavy inclination of using internet slangs such as jargons and abbreviations. For example, the words \emph{MRI}, \emph{CT} and \emph{OMG} are some of the short forms or abbreviations used in the tweets above. Tweet ``c'' consists of the word \emph{bypass}, which is a medical term for a surgery or a heart surgery. Third, there are numerous spelling errors with words such as \emph{loooowest} and \emph{healht} spelled incorrectly. Such examples also point out that individual social media posts or tweets in this case are usually very short or sparse. Table~\ref{social-text} highlights some of common issues found in social media text. Such issues necessitate proper cleansing and curation of data for robust analytics.

Ineffective or lack of robust cleansing and curation of such data may lead to the following implications:

\bemph{Faulty Decisions}
Without a robust cleansing and curation, social raw data fed into for deeper analytics is not fit for use~\cite{CurationAPI}. For example, consider an analyst who wants to classify tweets (from Twitter) related to doctors to ascertain general feedback as positive or negative. An analyst would rely on classifier or other analytical tools which would computationally parse the tweet and tag them with labels such as \emph{doctor} or \emph{other}. Tweets illustrated in Figure~\ref{fig:social-motivations}, would both be classified into class \emph{doctor}, however, this would be inaccurate for tweet (A).

The above example highlights that the lack of proper cleansing and curation leads to relevant data points not picked up or assigned incorrectly. In short, without an effective cleansing and curation, decision making can be severely compromised due to incorrect judgements.

\bemph{High costs}
Decision makers in organisations rely on data analytics to gauge customer sentiments and thereby improve marketing and other strategies. Social media data due to its popularity forms an key part of this analytics. Impacts of poor quality data on organisations has been widely studied; which stipulated that reliance on bad data can have adverse affects~\cite{haug2011costs,redman1998impact}. Such adverse affects could manifest in forms of dissatisfaction of customers, loss of business and credibility; all of which lead to higher costs to run the business in longer term. For example, misspellings in customers records (e.g., name and address) often lead to delivery errors in mail or products leading to higher maintenance costs for systems.

To summarize, we can compare raw data with a raw material such as iron ore. There is a limited use of raw iron; but once it is cleaned and alloyed into steel, it becomes useful for construction and other industries. Similarly curation offers a solution to transform raw data into something useful for analytics process \cite{beheshti2017automating}.

\section{Contributions}
\label{sec:intro_contributions}

To address the above mentioned challenges, in this thesis we propose,  an extensible social data curation pipeline for transforming raw social data into contextualized data and knowledge~\cite{beheshti2018crowdcorrect}. Two main phases  include:
\begin{inparaitem}[(i)]

	\item \textbf{Automated feature extraction and correction} - We design and implement micro-services to extract features such as keywords from a corpus of tweet data and automatically perform major data cleansing tasks on extracted keywords. \\
		\item \textbf{Crowdsourced correction} - We  extend our approach, to use the crowd inputs to further cleanse data which could not be corrected in the earlier step. In order to achieve this, we take extracted features (e.g., keyword) from the earlier step and automatically generate micro-tasks with possible options for the user to choose from. These micro-tasks are presented to users within a simple web interface. Our micro-tasks generation service uses external knowledge bases such as Bing spell check\footnote{https://azure.microsoft.com/en-au/services/cognitive-services/spell-check/} to suggest possible answers.
		
Further, we select a corpus of tweets; perform the automated and crowd correction step and then use a classifier to measure its accuracy. We compare the results against a similar classification without using our approach.
	\end{inparaitem}

\subsection{CrowdCorrect Curation Pipeline}
\label{sub:intro_crowdcorrect}

We propose CrowdCorrect as a social data curation pipeline consisting of several steps. The first step covers automatic feature extraction (e.g., keywords and named entities) and correction. We focus on three types of textual issues found in social media,  namely:

\begin{compactenum}
\item{\textit{Misspellings}} - we provide services to correct the spelling
\item{\textit{Abbreviations}} - we provide services to replace to full form, e.g., Aus. to Australia
\item{\textit{Jargons}} - we provide services to normalize with a more standard form, e.g., replace cardiologist with doctor
\end{compactenum}
 The automatic correction relies on external knowledge sources to identify the best possible match. In the second step, we design micro-tasks and use the wisdom of the crowd to identify and correct information items that could not be corrected in the first step. In this step, the micro-tasks are automatically generated using extracted features and possible correction suggestions are chosen using external knowledge sources. For example, we pick an extracted keyword such as \emph{helht} and use a spell check service to provide suggestions. We aggregate the results after running an experiment with crowd users and select the result with highest votes.
 The micro-task is then presented to a crowd user with a simple option to select the right suggestion as per their choice. We then aggregate the answers to avoid any user bias. CrowdCorrect is offered as an open source project, that is publicly available
on GitHub\footnote{https://github.com/unsw-cse-soc/CrowdCorrect}. Both the contributions are also further discussed below.

\subsection{Automated Feature Extraction and Correction}
\label{sub:intro_afe}
We implement a set of micro-services to automatically extract features and correct raw social data. These services will extract:

\begin{compactenum}
\item \textit{Lexical features} - such as keywords
\item \textit{Natural-Language features} - such as named-entities (e.g., person, product etc), part-of-speech (e.g., verb, noun etc.)
\item \textit{Time and Location features} - mentions of location or time within the data
\end{compactenum}

We then design and implement services to use the extracted keywords in the previous step and to identify and correct misspellings, abbreviations and jargons (i.e., special words or expressions used by a profession or group that are difficult for others to understand).
These services leverage external knowledge bases and services namely Bing\footnote{https://azure.microsoft.com/en-au/services/cognitive-services/spell-check/
}( for misspellings), Cortical\footnote{http://www.cortical.io/}(for synonyms) and STAND4\footnote{http://www.abbreviations.com/abbr api.php}(for abbreviations). The result of this step will be an annotated dataset which contain the cleaned and corrected raw data.

\subsection{Crowdsourced Correction}
\label{sub:intro_crowdsourcing}
Further we design and implement micro-tasks and use the wisdom of the crowd to identify and correct information items that could not be corrected in the first step. The micro-tasks are automatically generated using extracted keywords and possible correction suggestions are sourced from above mentioned knowledge bases and services. Crowd users are shown a generated micro-task in a Web browser to select the correct suggestion. Several rules are used to ensure maximum coverage of the feature dataset. These rules govern which feature item would be picked for the ``next'' microtask based on number of answers given by existing users. The output of this step is crowd corrected data.

\section{Thesis Organization}
\label{sec:intro_thesis_organization}

This thesis is organized as follows: In Chapter~\ref{chp:related_work}, we
present the background and state of the art in Social data curation, crowd-sourcing and crowd-sourced curation.
In Chapter~\ref{chp:apibase}, we present details about the design and
implementation for \emph{CrowdCorrect}.
In Chapter~\ref{chp:imp_evl}, we present details about the design and
implementation of the CrowdCorrect platform along with the experimental evaluation.
Finally in Chapter~\ref{chp:conclusion}, we provide concluding remarks of this
thesis and possible future work.
\chapter{Background and State of the Art}
\label{chp:related_work}

In this chapter, we start with a brief discussion on social media data and their issues. Following on further, we discuss what data curation and data cleansing are. We look at various techniques for cleansing and curating social media data, along with their limitations. Then, we examine the concepts of crowdsourcing with popular examples. Finally, we examine the use of crowdsourcing for social media data curation along with its limitations.

\section{Social Media Data}
\label{social-media}

Social networks or microblogging sites started appearing in public domain as early as 2003~\cite{king2015use}; when a site called MySpace\footnote{https://myspace.com/} was launched. By design, social network sites empowered their users to build relationships, communities, popularity, express opinions and concerns amongst other social benefits~\cite{van2013understanding,CrowdCorrect,MuhCrowdsourcing}. This drove popularity of social media sites~\cite{van2013understanding}; which continued to rise in the last decade with sites such as Twitter\footnote{http://twitter.com}, Facebook\footnote{http://facebook.com} and Instagram\footnote{https://www.instagram.com} having record number of users across the continents\footnote{https://www.smartinsights.com/social-media-marketing/social-media-strategy/new-global-social-media-research/}. Governments and organizations have also jumped onboard to examine their policies~\cite{ProcessAnalytics,hagen2018government,bertot2012impact,Galaxy,ProcessAtlas,BPM,FPSPARQL,CaseWalls}, develop products and guage sentiments~\cite{jeong2017social}, marketing strategies~\cite{tuten2017social,SocialTrust,BilalTrust,CNR,SETTRUST} and so on. As such, information shared online nowadays is predominately user generated content~\cite{figueiredo2014dynamics,GOLAP,WISEolap,AdaptiveRule,iSheets,iCOP}. Table~\ref{fig:social-sources} illustrates statistics of some of the popular sites as of 2018.

\begin{table}
\centering
\resizebox{\textwidth}{!}{
 \begin{tabular}{||c c c c ||}
 \hline
  & Twitter & Facebook & Instagram  \\
 \hline\hline
 Purpose & Micro-blogging & Social Networking & Social media sharing \\
 \hline
 Active Users & 328 million & 2.1 billion & 700 million  \\
 \hline
 Daily data Stats & 500 million tweets & 300 million photos & 95 million posts \\
 \hline
 Finer Stats (per second) & 6000 tweets &  500,000 links & ~4500 photos\\[1ex]
 \hline
\end{tabular}}
\caption{2018 Stats of Popular Social Media Sites}
\small \textbf{source}: https://www.leveragestl.com/social-media-infographic
\label{fig:social-sources}
\end{table}

\begin{figure}
\centering
\includegraphics[width=0.9\linewidth]{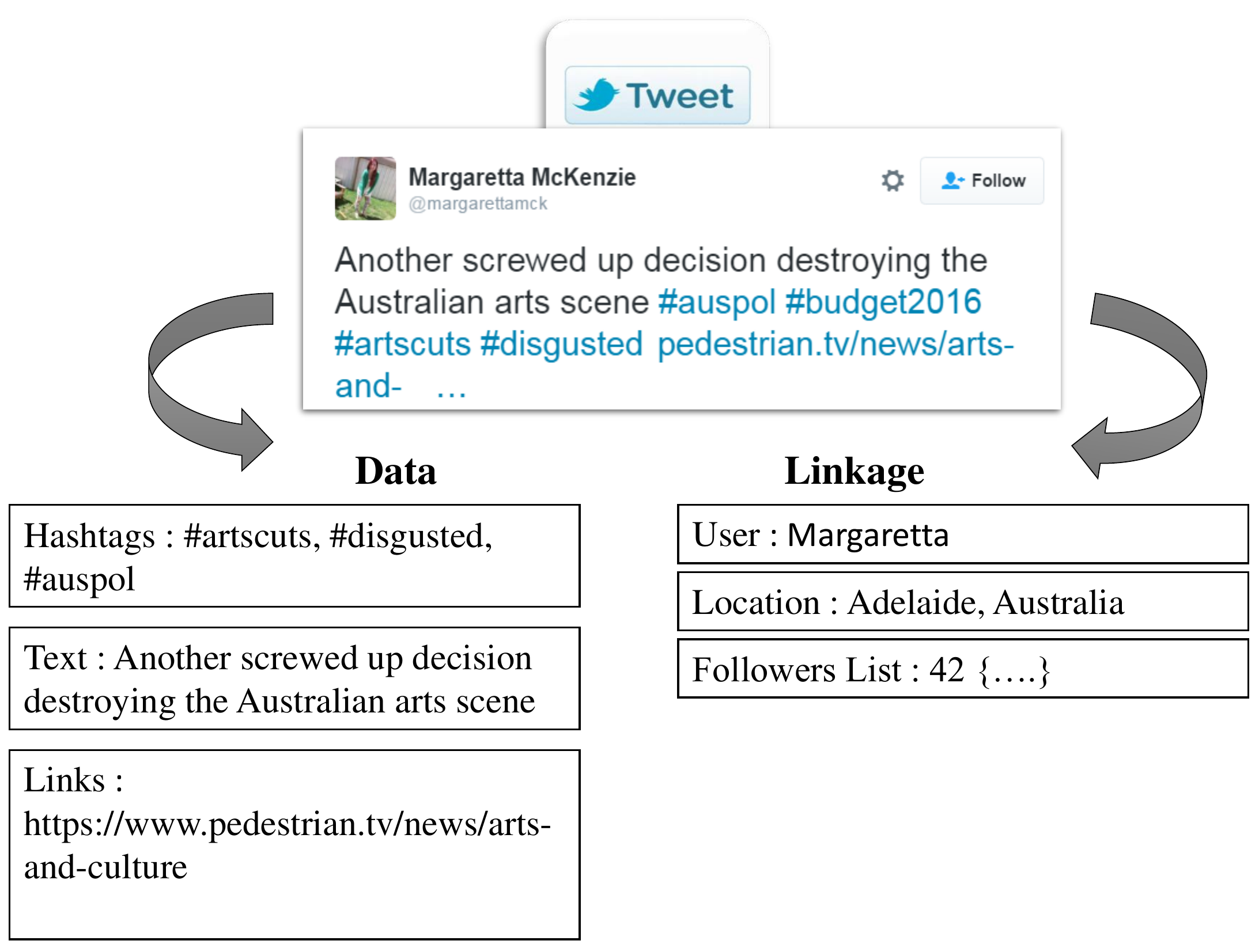}
\caption{Parts of tweet broken down.}\label{fig:pipeline}
\end{figure}

User content generated on social network sites along with linkage data~\cite{aggarwal_introduction_2011} (e.g., user information, friends, followers and location) is collectively referred as social media data. This data can be classified as \emph{open data}, since it is available publicly and can be queried~\cite{doi:10.1080/15228835.2012.743797}. Data from each social channel (e.g., Twitter and Facebook) can be broken down into individual messages or more commonly known as posts or blogs. As an example, each individual post on Twitter is termed as a tweet. Further, each post can contain several smaller artefacts such as text, media and hyperlinks. For illustration, take a look at Figure~\ref{fig:pipeline} for content and linkage data which can be extracted from a tweet on Twitter. Sometimes sites place limitations on words a post can contain (e.g., 140 characters for a tweet on Twitter); making it sparse. On any given day, there are millions of posts on popular social channels on diverse range of topics imaginable, therefore, making it a gold mine for information.

\emph{Velocity} and \emph{Volume} are well-known challenges in analysing social data~\cite{cambria2013big}; due to the size or load of data. This stems both from the popularity as well as round the clock availibility of social channels on Web and mobile. In addition, data collected is also often of either semi-structured (e.g., JSON format for a tweet) or unstructured (e.g., text) variety. However, it is also the quality of data, that poses significant challenges when making sense of it. As noted by a Eisenstein J.~\cite{eisenstein2013bad}, social medium contains user content which \emph{defies expectations about vocabulary, spelling, reliability and syntax}. For example, in part-of-speech tagging experiment, the Stanford tagger\footnote{https://nlp.stanford.edu/software/tagger.shtml} falls well below in accuracy when posed with Twitter content~\cite{gimpel2010part}. As such several studies such as~\cite{eisenstein2013bad}, have pointed out that state-of-the-art natural language processing (NLP) systems perform significantly worse on social media text.

A lot of research work has looked at techniques for domain specific approaches to harness social data. Examples of such work include disaster management~\cite{alexander2014social} and engagement for airlines operators~\cite{leung2013attracting}. Other approaches have focussed on specific attributes (such as \textit{like} feature on Facebook and \textit{re-tweet} on Twitter) on social sites while some research works have studied the popularity of the social media sites as well. Examples of such include news propagation ability in
Twitter~\cite{kwak2010twitter}, benefits of Facebook friends~\cite{ellison2007benefits}, recommendation system of 
YouTube~\cite{davidson2010youtube} and combining blogs with networking aspects of Tumblr~\cite{chang2014tumblr}. Finally, challenges in performing social media analytics such as data volume and quality have been pointed out in some research works such as in~\cite{stieglitz2018social,beheshti2017coredb,CoreKG}. Such works usually fall short of any specific guidelines to solve the issues.

Focussing on Twitter, there is large number of work presenting mechanisms to capture, store, query and analyze
Twitter data~\cite{goonetilleke2014twitter}. These works focus on understanding various aspects of Twitter data, including the temporal behaviour of tweets arriving in a Twitter~\cite{perera2010twitter}, user influence measurement in Twitter~\cite{cha2010measuring}, measuring message propagation in Twitter~\cite{ye2010measuring} and sentiment analysis of Twitter audiences~\cite{bae2012sentiment}.

The above mentioned research work is closely related, in that they focus on social media, but not directly related to our research problem. Our work is more  closely related to improving the quality of the social data via a robust curation before fed into for deeper analytics. The closest approaches to our work is in dealing with noisy text~\cite{baldwin_han_nodate,subramaniam_survey_2009}.  We further look at related work in data quality in Section~\ref{data_cleansing}.

\section{Data Curation}
\label{data-curation-rl}

In this section, we discuss the background and importance of data curation. Further on, we also look at related work and techniques for curating social media data.

Understanding and analysing data is considered as a vital capability for critical decision making in governments and
organizations~\cite{DataSynapse,CurationAPI,iProcess,CurationAPIs,CDCR,CDCRTech,MuhRating2,MuhRating1,Tagging}. Any issues with raw data introduce significant challenges in extracting actionable intelligence. Such issues as discussed in the previous sections, include noise (quality related issues), dealing with different data types (from structured to unstructured) amongst the volume and velocity of data. It is therefore, important to transfrom the raw data into contextualized data and knowledge; which analytics tools and end-users can consume. There are further rationale for a transformation of raw data for proper selection and preservation as discussed by
Lord et al.~\cite{lord_data_2004}. Given the nature of social media data (as discussed earlier), it is an ideal candidate for such a transformation.

\textbf{Data curation} is a process that takes raw data as an input and produces curated or contextualized data and knowledge; which can then be consumed for deeper analytics~\cite{DataSynapse,palmer2013foundations,TPM,AminProv}. Simply put in~\cite{cragin2007educational}, ``Data curation is the active and on-going management of data through its lifecycle of interest and usefulness; curation activities enable data discovery and retrieval, maintain quality, add value, and provide for re-use over time''. As such, the curation process abstracts and adds value to the data thereby making it useful for users engaging in analysis and data discovery.  In order to transform the raw data into contextualized data and knowledge; a curation process typically consists of a number of iterative activities, which is discussed in Section ~\ref{DatacurationProcess}.

The term curation, in the past commonly referred to library and museum professionals~\cite{moore1994museum}. The curators and their curation activities formed the backbone of musuem or library management. The skills and knowledge of the staff - the curators; added value to physical objects so as to provide context and history for their research and learning. In a way, data curation was a term formed to explicitly transfer curation guidelines and techniques as used by museum and library professionals on physical objects to
data~\cite{beagrie2008digital}.

Related techniques to data curation include ETL (Extract, Transform and Load) systems\footnote{https://www.informatica.com/au/}, entity deduplication~\cite{chaudhuri2005robust} and various other data integration systems such as schema
integration~\cite{chiticariu2007semi,rahm2001survey}, graph modeling and processing~\cite{DistributedGraph,DREAM,GraphSurvey} and federation of data~\cite{chen2005data}. Such systems are distinct from curation in that curation views transformation of raw data and the curation sub-tasks in a wholesome manner~\cite{stonebraker_data_nodate}. The goal and focus of such tools is not particularly on building a scalable curation~pipeline.

\subsection{Data Curation Activities}
\label{DatacurationProcess}

\begin{figure}[t]
\centering
\includegraphics[width=\textwidth,height=\textheight,keepaspectratio]{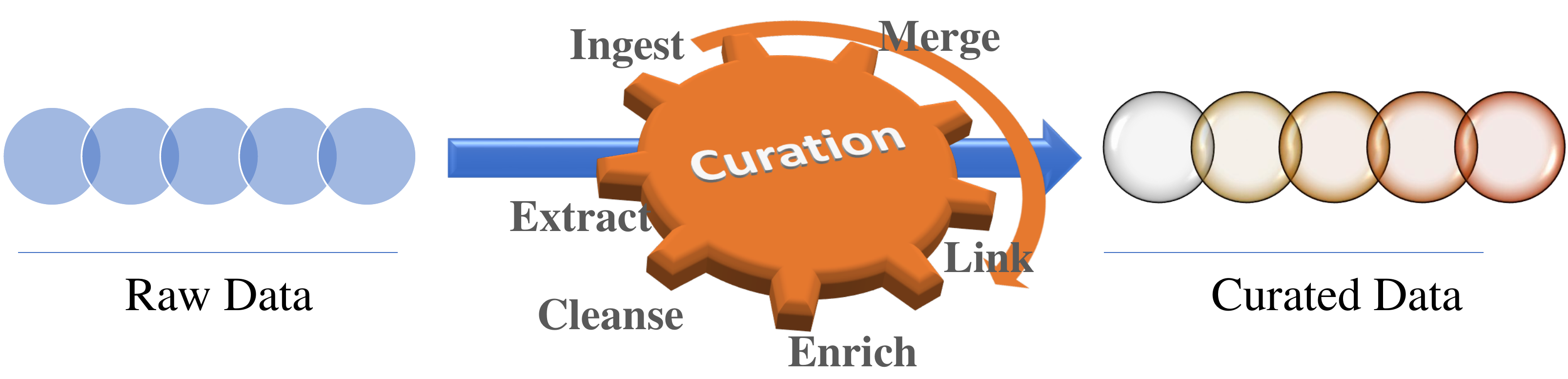}
\caption{Activities in a curation process.}\label{fig:curation-process}
\end{figure}

Data curation usually consists of a pipeline of iterative activities, techniques and algorithms~\cite{CurationAPI}.
Figure~\ref{fig:curation-process} highlights some of the major activities within a curation pipeline.
These activities include:

\begin{enumerate}
  \item \textbf{Ingest}: Identification and extraction of data and knowledge e.g., from a data source such as a database~\cite{grover2015data} or using human~\cite{sadeghi2015viske};
  \item \textbf{Cleanse}: process to improve the quality of data, e.g., identify and remove unwanted items from data~\cite{krishnan2016towards}. Cleansing improves data quality (discussed in depth in Section ~\ref{data_cleansing}); while linking and enriching add value to the data;
  \item \textbf{Link}: process to link data with other relevant data items, e.g., entity linking~\cite{shen2015entity,elmagarmid2007duplicate};
  \item \textbf{Enrich}: Use internal as well as external sources to enrich the data, e.g., use knowledge bases such as Wikipedia\footnote{https://en.wikipedia.org}~\cite{troncy2016linking};
  \item \textbf{Merge}: Identify and merge data as relevant, e.g., merging of data streams~\cite{eul1996method};
  \item \textbf{Maintain}: Preserve and make data available as required, e.g., store data in formats to promote re-use~\cite{rundensteiner2000maintaining,chaudhuri1997overview}.
\end{enumerate}

\bemph{Simple Curation Pipeline Example}
Let's consider an example from Twitter as a simple curation process. Say we want to perform analytics on English language tweets related to today's news in Sydney, Australia. It is possible to Ingest tweets using Twitter's API\footnote{https://developer.twitter.com/en/docs.html} and store them inside a relational database such as SQL Server\footnote{https://www.microsoft.com/en-us/sql-server/sql-server-2017}. Then, we can cleanse any tweets which are not in English; using a tool such as LingPipe\footnote{http://alias-i.com/lingpipe/demos/tutorial/langid/read-me.html}. In this cleansing step, we can also automatically correct any misspellings using any off-the-shelf spellcheck tool such as Bing Spell Check API\footnote{https://azure.microsoft.com/en-us/services/cognitive-services/spell-check}.

\begin{figure}[t]
\centering
\includegraphics[width=\textwidth,height=\textheight,keepaspectratio]{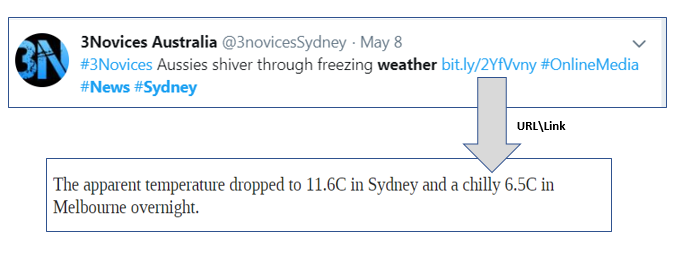}
\caption{Example of extracting and annotation of Twitter data.}\label{fig:curation-sample}
\end{figure}

 Further, we can add value, by annotating the tweets with additional information contained in the URLs or links inside the tweets. Refer to Figure~\ref{fig:curation-sample}, by extracting the content on the link within the sample tweet, we could annotate the tweet with additional information such as ``\textit{11.6c}'' cold expected in Sydney. We store the annotated tweets inside the database as a \emph{curated set} and then perform analytics on them. For example, we could classify the various news items into sports, weather and  politics. Then rank them in order of social media popularity, i.e., number of tweets.

\subsection{Data Curation Approaches and Frameworks}
\label{dc-app-frame}
Data Curation is an umbrella term of activities which is often combined into one or several approaches. This section lists some of the popular curation approaches. Further, we look at  curation platforms and Application Programmers Interface (API) discussed in literature.

\subsubsection{Approaches}
The curation activities discussed in the previous section are usually combined inside a curation approach depending on the end goal (e.g., analytics, curate content only and enrich).
These approaches to data curation along with identified research works are shown in Figure~\ref{curation-approaches}.

\begin{table}[ht!]
\begin{tabular}{|p{5cm}|p{5cm}|p{2cm}|}
\hline
\textbf{Approach}       & \textbf{Technique}                                                                          & \textbf{References}   \\ \hline
Collaboration Platforms & Curate content such as news, blogs using an online platform e.g., Storify\footnote{https://storify.com/}, Wikipedia\footnote{https://wikipedia.org}.   &    \cite{bruns201518}, \cite{yakel2007digital}      \\ \hline
Curation at source      & Integrate lightweight curation activities in other workflows.                                &   \cite{curry2010role}, \cite{hedges2012sheer}    \\ \hline
Master data management  & Create and maintain single source of data with curation activities performed on it.          &    \cite{otto2010enterprise}, \cite{morris2005managing}    \\ \hline
Crowdsourcing           & Utilize the collective wisdom of crowds to perform intensive or simpler curation activities. & \cite{doan2011crowdsourcing}, \cite{porcello2013crowdsourcing}  \\ \hline
Curation At Scale & End-to-end data curation pipeline. & \cite{stonebraker_data_nodate} \\ \hline
\end{tabular}
\caption{Approaches to Curation.}
\label{curation-approaches}
\end{table}

Colloboration platforms allow content to be aggregated or extracted from various sources and curated collectively by users. Platforms such as Storify\footnote{https://storify.com/} are prime examples of such platforms. These platforms rely on user inputs and their motivations to collaborate. In other words, due to manual nature of curation tasks, the process is often time consuming. Also the end-goal here is to develop a story from already published content (Twitter and news articles) and not on improving the quality of the underlying data.

 On the other hand, master data management approach focusses on creating a single source of curated data for an enterprise. This might sound ideal but challenging given the amount of data (along with number of sources) in any organisation's nowadays along with the need to create a uniform model across it. Further, mining one single source of data is challenging and there are quality 
 trade-off's~\cite{silvola2011managing}.

  Curation activities can also be integrated and combined into other workflow activites for example embedding capture and curation as a part of researcher's working practices~\cite{hedges2013digital}. While such an approach may help customisation; this may lead to bespoke and non-standardization leading to increased maintenance.

   Curation at scale implies building an end-to-end curation pipeline for scalability, automation and quality. Finally, data curation can be a resource intensive and complex task, where it's beyond the capacity of a single individual. Using crowdsourcing platforms such as Amazon Mechanical Turk\footnote{https://www.mturk.com/}, certain well defined curation tasks such as cleansing can be outsourced to a crowd of users~\cite{freitas_big_2016}. Although, the crowd-sourced approach can effectively use the wisdom of the crowd users; the participation and motivation of the users can introduce challenges such as longer times, user selection and biased opinions.

   Our research focusses on improving the quality of the social media data by a robust cleansing process within an extensible curation pipeline. Our data curation approach is different as compared to the ones discussed above; in that we combine the automated and scalable nature of curation at scale; along with crowdsourcing to achieve our goals.

\subsubsection{Curation Platforms and APIs}

Curation platforms provide a curation pipeline with a focus on all or a particular curation activity (e.g., linking). Below we discuss some of the curation platforms.

DataTamer~\cite{stonebraker_data_nodate} is an end-to-end curation system for integrating and transforming multiple data sources into a single predefined data structure for further reuse. The system uses machine learning algorithms to inspect a data source and then automatically extract entities, perform deduplication, transformation and mapping. Further, a human can intervene and specify transformations and mapping manually via a user interface. Then an expert or domain expert validates the data transformation. However, DataTamer is not designed to perform any quality checks on the data itself; as is the case with our research. Also DataTamer system is not geared towards the largely unstructured nature of social media data.

Another system ZenCrowd~\cite{demartini2012zencrowd} proposes a curation process with a focus on linking entities in text to an external knowledge base. The system works by automatically extracting limited set of features (e.g., persons, entities and organisations) from an HTML page text. Then it uses an algorithmic matcher to extract more information about each extracted feature from linked open data cloud\footnote{https://lod-cloud.net/}. The results of the algorithmic matcher are scored using a probablistic method. Further, low scoring results are passed onto a crowd task module to automatically create a crowd task; which is posted on a crowdsourcing platform. While the results show higher precision; there is uncertainty when entities would have any number of textual imperfections (e.g., misspellings and abbreviations) as prevalent in social media.

Several other similar approaches to assist a curation process have been discussed in literature. Kurator~\cite{dou2012kurator} is a curation workflow for aiding curation for scientific data; although limited to spreadsheet data. Many commercial tools such as Dremio\footnote{https://www.dremio.com/} and Snowflake\footnote{https://www.snowflake.com/} have also sprung out; which can be leveraged to build a custom curation process.

In addition to curation systems discussed above; Beheshti et al.~\cite{CurationAPI} proposed a set of basic data curation APIs. These APIs are exposed as RESTful services such that they can be used by researchers and developers alike. The services cover extraction, linking and classification of raw~(open)~data.

\section{Data Cleansing in Curation}
\label{data_cleansing}
Data cleansing is an important phase in the data curation process, as we discussed earlier. Considering the theme of our research work, we now discuss the background on data cleansing. Further we look at issues and existing techniques for cleansing social media like data.

\subsection{Data Quality}
\label{data-quality}
In order to better understand the data cleansing process, it is important to first understand the concept of data quality. One popular way to understand data quality is to comprehend
its ``fitness for use''~\cite{watts2009data}. Similar definitions exist, for example researchers in~\cite{wang1996beyond}, define data quality as data that are fit for use by data consumers. Also in data quality literature, data  is associated with several dimensions that imply overall quality. For example, in another such research work, data is broken into several dimensions such as timeliness, accuracy, completeness and consistency to guage its quality~\cite{lederman2003meeting}. Researchers corroborate that accuracy is straightforward to evaluate as its merely comparing the correct value versus the observed value. A further argument is that timeliness and completeness are also relatively straightforward to evaluate. Consistency is viewed as slightly more complex as it relies on ongoing comparison of other dimensions. Other research works have added more dimensions such as interpretability and 
accessibility~\cite{wang1995toward}. Yet another research work~\cite{wang1996beyond}, classifies data into categories such as intrinsic, contextual, representational and accessibility, each having a set of dimensions.

\begin{figure}[ht!]
\centering
\includegraphics[height=3.5in,keepaspectratio]{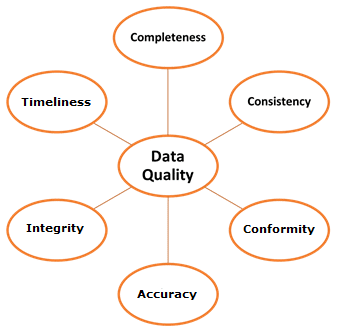}
\caption[Caption for LOF]{Dimensions for Data Qaulity\protect\footnotemark.}
\label{fig:crowd-content}
\end{figure}
\footnotetext{https://smartbridge.com/data-done-right-6-dimensions-of-data-quality/}
Data with poor quality is also referred to as dirty or bad data \cite{wang1996beyond}. The impacts of poor quality data are serious. For example, for businesses, this can have negative consequences such as increased costs, inaccurate decisions leading to unsatisfied customers. 
A recent study~\cite{ballou2003special} has shown that bad quality data can lead to not only economical but also social consequences for organisations. As per estimates discussed in~\cite{khayyat2015bigdansing}, around 25\% of data in organisations is dirty. This includes both structured (for example an entity stored in a relational database) and unstructured( for example text and emails) Data quality in largely unstructured online user generated content, as in social media sites (e.g., Twitter), is more averse. As noted earlier, Eisenstein et. 
al.~\cite{eisenstein2013bad} view social medium  content as one that defies expectations about vocabulary, spelling, reliability and syntax. We discussed common problems with social media data earlier which include slangs, non-standard text and grammar. Such linguistic noise are often couple with brevity (140 characters for tweet) making the quality of data poor.
In the next section, we look at data cleansing; an activity to improve data quality.

\subsection{Data Cleansing}
Data cleansing is a vital task to improve the quality and thus usefulness of data. From a process perspective, \bemph{data cleansing} is defined as enitirety of operations performed on existing data to remove anomolies such that the result set of data is accurate representation of the mini-world~\cite{muller2005problems}. An \textit{anamoly} is typically a data value which has an incorrect representation~\cite{muller2005problems}. Common anomalies are often inaccurate, duplicated or incomplete pieces of data. Anomalies usually arise due to erroneous inputs or measurements while collecting, inputting or maintaining data. In simpler words, cleansing makes the data fit for use by removing any uncertainties in data.

\section{Social Media Data Curation}

In social media curation, the focus is to transform raw data (unstructured to semi-structured) into curated data. Key challenges include; ingestion of continuous flowing social data, cleansing of the data due to its noisy nature as discussed earlier, linking and enriching the data for a given context.

Past research work highlighted the need for curating social media data as pointed out by Duh et. al. in~\cite{duh2012creating}; given social media's wider reach and acceptance. At the same time challenges in data collection, preparation and analysis have also been widely articulated~\cite{agarwal2010information}. Mining social media sites rely on parts of curation techniques for specific purposes. For example, taking some examples from Twitter related research there has been work done to understand the emotions in a 
tweet~\cite{roberts2012empatweet}, identify mentions of a drug in a
tweet~\cite{ginn2014mining} or detecting political opinions in tweets~\cite{maynard2011automatic}. Another research study highlighted the
need for curating the Tweets but did not provide a framework or methodology
to generate the contextualized version of a tweet~ \cite{duh2012creating}.

One of the closer related work is lexical normalisation of social media text~\cite{han2013lexical}. The proposed approach uses a classifier to target out-of-vocabulary words and normalises lexically similar words. This works well in isolation for a subset of noisy text issues; without an aim to contextualise the data for analytics.

\section{Crowdsourcing for Curation}
\label{crowdd}
While technology continues to evolve at rapid pace; there are many problems where human intelligence and interpretation is more effective. As an example, consider a simple task of tagging images with types of animals such as dog, cat or horse. This is relatively hard for a computer program to analyse given the different physical characteristics of various animals. On the other hand, such a task is relatively easier for humans. Employing a labour workforce to accomplish such tasks is both time-consuming and expensive. With the rapid advancement and scalability of Web technologies, outsourcing tasks to a crowd of users has become popular~\cite{brabham_crowdsourcing_2008}. This has led to keen interest in the research community on areas such crowdsourcing concepts, effectiveness, crowd selection, crowd motivation, tasks design and practical use cases.

\subsection{Crowdsourcing Concepts}

The term crowdsourcing was coined in 2006 in~\cite{howe2006rise} as ``taking  a function once performed by employees and outsourcing it to an undefined (and generally large)  network of  people in the form  of an  open  call''. The underlying principle asserted 
in~\cite{surowiecki2007wisdom}, is that the collective wisdom of a large group of people produce superior results than an individual. Today's growth in Web and mobile technologies have created an atmosphere to tap into distributed large groups of people at 
scale~\cite{kietzmann2011social}. This has also led to the original definition of crowdsourcing in~\cite{howe2006rise} rather obsolote; with crowdsourcing campaigns maturing to target specific crowds, availability of crowdsourcing API (application programming interfaces) and combination of machine and human inputs~\cite{kietzmann2017}.

Popular examples of crowdsourcing are Wikipedia\footnote{https://www.wikipedia.org/} and Threadless\footnote{https://www.threadless.com/}. Wikipedia allows volunteers around the world to create and edit content. Threadless allows a community of users to select and create t-shirt designs for an incentive. Further platforms such as Amazon Mechanical Turk\footnote{https://www.mturk.com/
} and UpWork\footnote{http://www.upwork.com} have allowed organisations to rapidly create and deploy crowd tasks at scale. Another crowdsourcing platform Figure-Eight (formerly known as CrowdFlower)\footnote{https://www.figure-eight.com/} provides an  ability to annotate unstructred data with crowd judgements for feeding as training data to machine learning programs. Several research studies have applied crowdsourcing to solve problems such as assembling dictionaries~\cite{lanxon2011oxford}, outer space 
mapping~\cite{mclaughlin2014image} and aiding in disaster  relief situations~\cite{zook2010volunteered,rogstadius2013crisistracker}.

Before any form of crowdsourcing can take place; both the problem and anticipated crowd inputs must be clearly defined. Several 
studies~\cite{prpic_how_2015,prpic2016crowd,kietzmann2017} have categorized content or inputs from crowd as either objective or subjective; and contributions or responses from users to be either aggregated or filtered. An illustration from~\cite{prpic_how_2015} is shown in Figure~\ref{fig:crowd-content}. The four types of crowdsourcing are shown with an example for each. Idea and solution crowdsourcing contributions can be termed macro-tasking; whereas crowd-voting and micro-tasks can be thought as micro-tasking due to the level of granularity~\cite{garcia-molina_challenges_nodate}. Responses from micro-tasks based crowdsourcing are usually aggregated, e.g., aggregate the total votes from a poll. On the other hand, contributions from macro-tasks based crowdsourcing are usually selected or filtered as required.

\begin{figure}[t]
\centering
\includegraphics[width=5.5in,keepaspectratio]{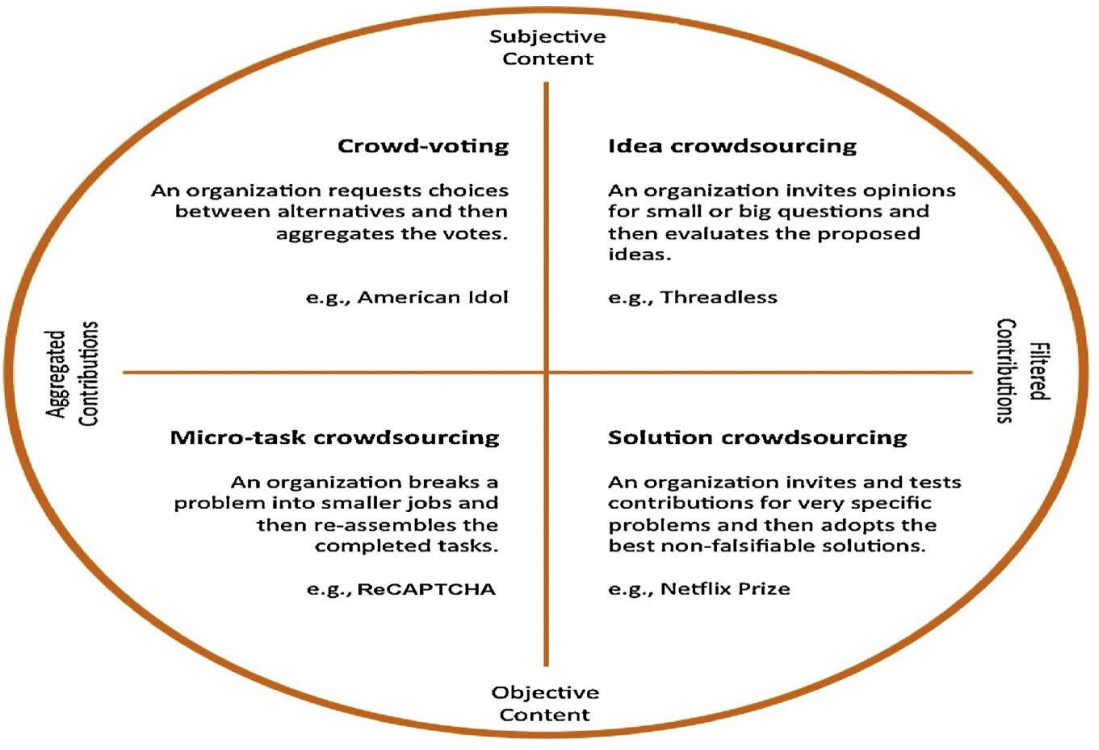}
\caption[]{Ilustration of crowd sourcing types\protect\footnotemark.}
\label{fig:crowd-content}
\end{figure}
\footnotetext{ https://www.slideshare.net/IanMcCarthy/how-to-work-a-crowd-developing-crowd-capital-through-crowdsourcing-44425140}

\subsection{Crowd Participation and Crowdsourcing Effectiveness}

The success of crowdsourcing campaign depends on the performance of the crowds. Several recent studies such 
as~\cite{garcia-molina_challenges_nodate} have highlighted that crowd of users can be slow, give wrong answers or opinions and also use the platform to spam without doing any work. There is a tradeoff with uncertainity when dealing with contributions from the crowd. They propose to have proper worker evaluation techniques~\cite{garcia-molina_challenges_nodate} or compute consensus~\cite{hung2018computing} such as aggregating results to remove any undesired contributions. Further they highlight the importance of proper technical infrastructure setup for example an easily accessible web-based tool and tasks design so as to improve crowd efficiency and accuracy.

Researchers have also looked closely at what makes a crowd tick to understand the effectivess of crowdsourcing. Relying on a pool of potentially unknown crowd of people can be like a double edged sword instead of a trsuted employee. Such studies have brought forth issues like incentives~\cite{kaufmann2011more,rogstadius2011assessment}, gaining social capital~\cite{raban2008incentive}, game 
mechanics~\cite{cooper2010predicting} or general public good~\cite{kuznetsov2006motivations} plays a part in motivating a crowd user.

\subsection{Micro-task Design}

Our research is closely tied to creating microtasks for social media curation; therefore it is important to discuss studies pertaining to microtask design. As discussed earlier, microtasks are usually have low level of granularity with contributions that need to be aggregated for better results.

A recent study by Gaidraju et al.~\cite{gadiraju_human_2015} broke down microtasks in several types such as:
 Information finding, Verfication and Validation,  Interpretation and Analysis, Surveys and Content access.

Information finding tasks delegate process of searching to a crowd of users. As an example, ``find a hospital in West London''. Verification and validation or moderation tasks require a crowd user to validate a piece of information such as ``Is Italy a country ?''. In Interpretation tasks, crowd workers are often asked to use their mental skills such as ``Choose the best colour for Father's day''. Content access tasks require crowd users to view or access a content such as an advertisement.

The choice of the task type for a problem has a direct implication to the accuracy of results. Since microtasks are often undertaken by non-experts, they need to be simple to
process both mentally and logistically. They should not be too time-consuming nor
should they require a high degree of expertise or too much introductory training~\cite{vcibej2015role}. Several research 
studies~\cite{garcia-molina_challenges_nodate,rumshisky2011crowdsourcing} have proposed guidelines for defining input statement or problem, determining task type, task interface design and finding workers for best results. For interface design, the guidelines point to designing simple tasks with clear, short instructions to attract workers and reduce human errors. However at the same time they highlighted that interface design for crowdsourcing is similar to dark art and lot of more research is required to understand its impact on crowd performance and accuracy.

\subsection{Crowdsourcing for Curating Social Media Data}
\label{crowd-for-cur}

Crowdsourcing has been used to curate social media posts themselves. For example tools such as Storify\footnote{https://storify.com/} and Curated.by\footnote{https://www.curated.by} allow users to collect and curate tweets into stories making them easier to read.
Due to its inherent scalability, crowdsourcing can be used to leverage collective wisdom of a group of people for many data processing and curation tasks~\cite{garcia-molina_challenges_nodate}. Current limitations of machine learning algorithms and computer programs in tasks that are fairly easier for humans to process are also candidates for crowdsourcing~\cite{wang2012crowder}. For example crowdsourcing has been studied and quantified for extraction~\cite{heipke2010crowdsourcing},  collection of data~\cite{sigurdsson2016hollywood}, data cleansing and assessment~\cite{tong2014crowdcleaner,chu2015katara,hutchison_crowdsourcing_2013}, entity-resolution~\cite{wang2012crowder} and enrichment~\cite{dalvi2009web}. Such work relies on designing a crowd facing tool or interface and gathering contributions or answers from a crowd of users then aggregating the results.

To our knowledge a lot of such research highlighted above deals with curating and cleansing data which is essentially structured, i.e., for example in a relational database or non social media related.
The sparse, unstrcutured and noisy form of social media data coupled with limitations in natural language processing makes crowdsourcing an attractive preposition. There have been limited work or research looking at using crowdsourcing for curating social media data. One such work CrisisTracker~\cite{rogstadius2013crisistracker} extracts tweets from Twitter in real-time during a natural disaster. It then automatically detects localised events or stories based on clustering of tweet's data. Finally, the system uses a crowd of users to curate a story by ranking them. Here crowdsourcing is used for a limited purpose of ranking.

\section{Summary and Discussion}

Analytics of social media data is quite important and can be a vital priority and asset for organisations and government. This has been driven by large footprints of the social media channels such as Twitter. As we discussed in Section~\ref{social-media}, there has been prior research work on various facets of social media such as~\begin{inparaenum}[(i)]
\item \textbf{Domain specific} such as disaster management using Twitter,
\item \textbf{Feature specific} such as \textit{re-tweet} feature in Twitter or \textit{like} in Facebook,
\item \textbf{Popularity} of the social media sites,
\item \textbf{High level Analytics} that visualise trends in and social media such as popular topics
\end{inparaenum}.

One of the biggest challenge to perform accurate analytics of social media data is the poor data quality (Section~\ref{data-quality}) due to non-standardization of user inputs. These quality issues discussed in Section~\ref{social-media} include misspellings, grammatical errors and use of slangs. Further we saw how terms and words can imply different meanings when used in different contexts. Such linguistic problems of social media data often introduce more challenges in computational analysis~\cite{adedoyin-olowe_survey_nodate}.

Data curation (Section~\ref{data-curation-rl}) helps in transforming the raw data into contextualised data and knowledge, which can then be used for deeper analytics. There are several known approaches to curating data and prior research work has also looked at proposing curation platforms or services  (discussed in Section~\ref{dc-app-frame}). The key approaches are~\begin{inparaenum}[(i)]
\item \textbf{Collaboration platforms} which help collectively curate content from various sources;
\item \textbf{Master Data Management} where the purpose is to build one single model of data;
\item \textbf{Curation at Source} where curation is viewed as a part of other larger task;
\item \textbf{Curation at Scale} where the aim is to build an extensible end-to-end pipeline; and
\item \textbf{Crowdsourcing} that uses inputs from crowd users to help curate data
\end{inparaenum}.
 In addition to above, several platforms such as DataTamer~\cite{stonebraker_data_nodate} have been proposed which have limitations in dealing with low quality data.

Further, we then discussed (in Section~\ref{crowdd}) the background and benefits which Crowdsourcing brings along with important issues to address such as task design and user motivations. Many popular examples and platforms for crowdsourcing were illustrated. Following on, we discussed some applications where crowdsourcing has been used to curate social media data in Section~\ref{crowd-for-cur}.

\subsubsection{Challenges and Recommendation}

We have acknowledged that we need to transform the raw data into contextualized data and knowledge for carrying out robust analytics. Our key motivation is to help improve the underlying data quality, which is low in the social media channels. As an example, let us consider a simple curation pipeline to ingest, cleanse and enrich a corpus of few thousand tweets. To our knowledge, there are no existing off-the-shelf solutions that cater to this problem. There are however, several individual components such as a spell checker. Further, many existing curation approaches cater for structured data and are not geared towards the unstructured nature of social media. Various curation approaches are proposed in literature that are usually intended for specific purposes. We haven't come across an approach that leverages both automated curation along with the power of crowd-based curation into a single pipeline.

 Our research focusses on improving the quality of the social media data by a robust cleansing process within an extensible curation pipeline. Our approach is different in that we combine the automated and scalable nature of curation at scale; along with crowdsourcing to achieve our goals.
\chapter{CrowdCorrect}
\label{chp:apibase}

In this chapter, we discuss the need for curation pipeline for cleansing social media data. In particular, we discuss the pitfalls of certain standard cleansing approaches. Following on, we present an extensible data curation pipeline, CrowdConnect, to enable analysts to cleanse and prepare social data for reliable analytics. The  design of CrowdConnect includes microservices to ingest, extract and correct raw data. The correction services leverage automatic as well as crowd sourced approaches.

\section{Introduction} \label{sec:apibase_introduction}

Data cleansing or correction aims to improve the quality of data by removing errors and inconsistencies~\cite{rahm2000data}. As discussed in previous chapters, in the context of social media this is challenging due to high usage of slangs, abbreviations and acronyms. Data cleansing forms an integral part of the data curation activity. Before cleansing, raw data needs to selected, ingested and key features (e.g., keywords) needs to be extracted within a curation pipeline. We discuss related techniques in literature to social media data cleansing followed by our approach.

\subsection{Related Work}
 As a response to such unusual style and syntactical error prone nature of social media data text; the research community has looked at two major approaches namely, normalization and domain adaptation or contextualization~\cite{eisenstein2013bad}.

 Normalization approaches tend to find and replace non-standard words or terms with contextually correct ones. In other words, the idea is to fix or fit the data such that analytics tools can consume. A familiar example is of spelling checker 
 algorithms~\cite{hodge2003comparison}; which uses pattern matching and n-gram analysis to correct words. For example, the tweet \textit{``njoying at a bday''} is normalised to \textit{``enjoying at a birthday''}. Other examples of such approaches are machine 
 translation~\cite{aw2006phrase}, Twitter pre-processing approaches~\cite{clark2011text} and noisy channel 
 models~\cite{cook2009unsupervised}.

  Contextualization techniques works in reverse, i.e., making the tools smarter to adapt to bad data. These techniques apply nature language processing (NLP) algorithms like part-of-speech tagging~\cite{gimpel2010part,owoputi2013improved} and named entity recognition~\cite{finin2010annotating} to label and train the cleansing process. Essentially such approaches stem from a closely related field of noisy text analytics. The closest work in this category to our approach is the noisy-text project\footnote{https://noisy-text.github.io/norm-shared-task.html}. This research work is close in that it deals with quality issues in text in general; but does not look at wider issues in social media data.

Despite such work, curation and cleansing of social media text remains a challenge. The cleansing of social media data goes beyond a simple spell correction. The range of problems presented with out-of-vocabulary words, abbreviations, slangs, inconsistent grammar and the use of emoticons; make automated normalization or contextualization difficult if not impossible~\cite{clark2011text}. Normalization assumes that there is some direct mapping from out-of-vocabulary words to normal words. This can be misleading for social media data. For example, do we normalize the abbreviated slang word \textit{pat.} to \textit{patient} or something else. Further, some words such as \textit{howdy} have no direct mapping in English. Also, incorrect normalization can also result in semantic ambiguity~\cite{eisenstein2013bad}. For example how do we normalize the Twitter post, \textit{howdy baby}. Automated contextualization using parts-of-speech tagging also has many limitations. For example, Twitter data is composed of so many different styles and slangs with a lot of exceptions. The inherent presence of other non-standard textual items such as hashtags make tagging or named entity recognition difficult. To sum up, Einsentein et 
al.~\cite{eisenstein2013bad} illustrates that state-of-art Natural Language Processing (NLP) systems perform significantly worse on social media text.

Crowdsourcing has shown potential in problems which are relatively easier to solve for humans such as image labeling~\cite{he2013you} or annotationg parts of text~\cite{finin2010annotating}. Also, crowdsourcing can be leveraged to accomplish tasks on a global scale by rapidly mobilising large number of people~\cite{kittur2013future}. As an example, anyone with access to internet can perform micro-tasks using platforms such as Amazon Mechanical Turk\footnote{https://www.mturk.com/
} or Figure Eight\footnote{https://www.figure-eight.com/}. Social media services such as Twitter also have support for publishing simple tasks using Twitter Polls\footnote{https://help.twitter.com/en/using-twitter/twitter-polls
}. One could also put together a simple web-based interface and share micro-tasks with friends,  colleagues or anyone else. Crowdsourcing has already been used for collection of data~\cite{sigurdsson2016hollywood}, data cleansing and 
assessment~\cite{tong2014crowdcleaner,chu2015katara,hutchison_crowdsourcing_2013}, entity-resolution~\cite{wang2012crowder} and 
enrichment~\cite{dalvi2009web}.

\subsection{CrowdCorrect}
In order to address the challenges discussed above, we combine automated approaches with crowdsourcing approaches into an extensible curation and cleansing pipeline, CrowdCorrect. Our rationale is that a human should be able to identify and correct issues such as slang words given a clear well-defined task; relatively easily. As such, the cleansing of social media text can benefit from leveraging crowdbased approaches along with automated approaches. More specifically, this chapter discusses three phases that form the pipeline:

\begin{inparaenum}[(i)]

\item \textbf{Pre-processing: \textit{Ingestion and extraction}} techniques that leverage off-the-shelf tools and APIs to ingest raw data and extract features (e.g., keywords, named-entities etc.) on social media data;

	\item an \textbf{Automated curation: \textit{extraction and correction}} techniques that leverages external knowledge bases and services to automatically correct features on social media data;
	
		\item a \textbf{Crowdsourced curation: \textit{correction}} techniques that uses a crowd of users to identify and correct features which failed in the earlier step.\end{inparaenum}

\begin{figure}[h!]
\centering
\includegraphics[width=\textwidth,height=\textheight,keepaspectratio]{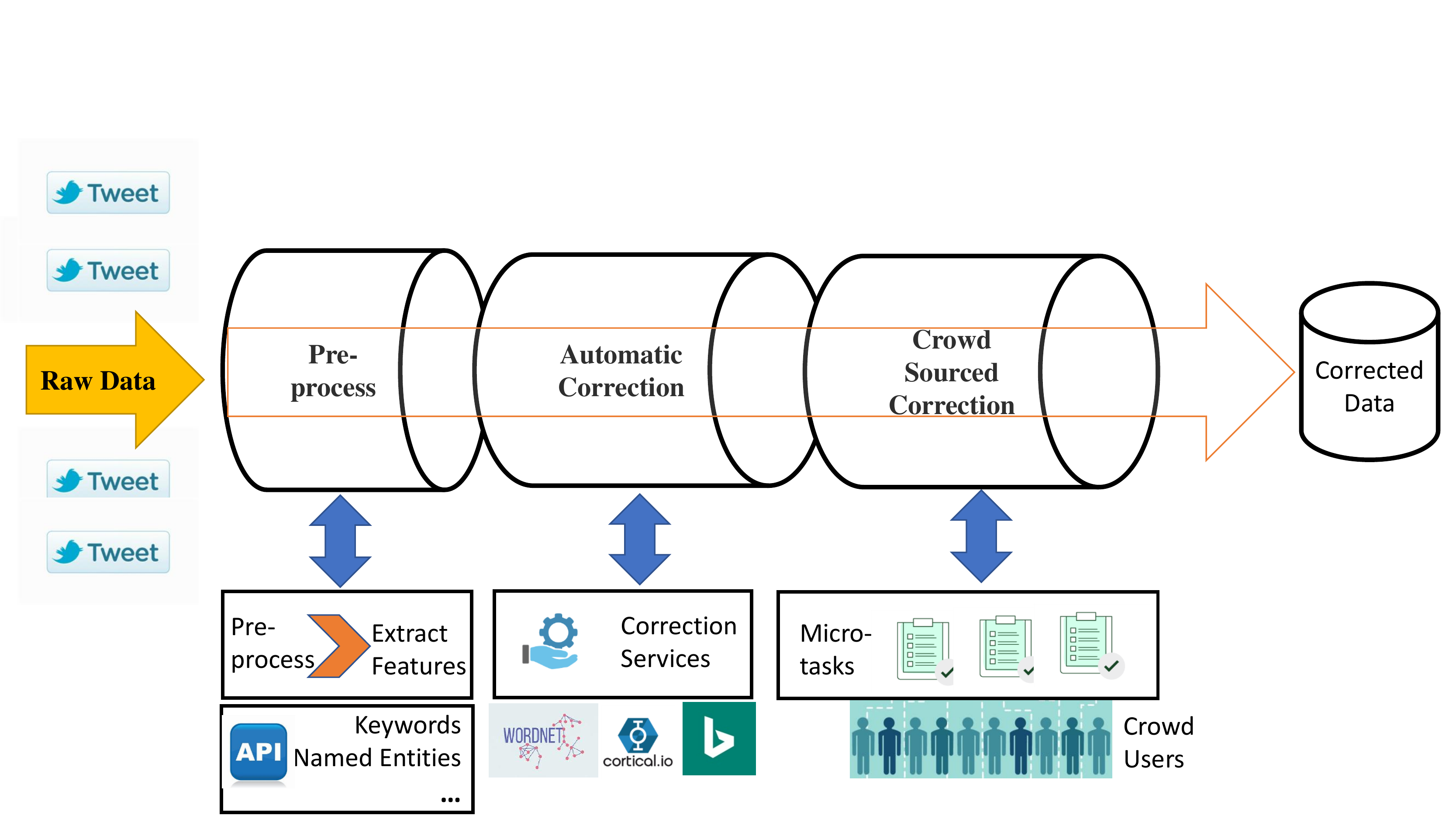}
\caption{CrowdCorrect Curation Pipeline.}\label{fig:crowd-correct}
\end{figure}

An overview of our curation pipeline is illustrated in Figure~\ref{fig:crowd-correct}, to enable analysts cleansing and curating social data and preparing for social media analytics. There are three steps to pre-process (ingest and extract), automatically correct and crowd-sourced correction. As an example, tweets are presented as raw inputs. The following sections discuss in detail about our contributions as shown in the illustration. We use and describe examples from Twitter as that is our social media site for research purposes.

\section{Pre-processing : Ingestion and Extraction}
\label{sec:apibase_ingestExtract}

This section presents an architectural overview of the ingestion of raw data and extraction of features from the data.

\begin{figure}[h!]
\centering
\includegraphics[width=\textwidth,height=\textheight,keepaspectratio]{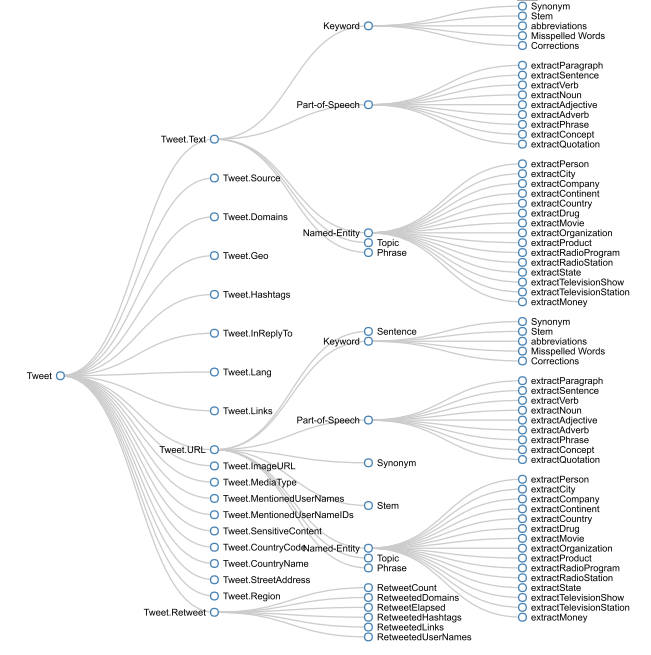}
\caption{An example from Twitter: Extraction services.}\label{fig:extraction}
\end{figure}

At first, we develop services to ingest data from social media channels such as Twitter (refer Section~\ref{sub:apibase_ingestion}). Ingestion takes the data and makes it available within our data store. Then, we perform extraction of features (e.g., keywords) using off-the-shelf extraction micro-services; developed previously within our research group. These services are outlined in the research paper by Beheshti et. al.~\cite{CurationAPI} and illustrated in Figure~\ref{fig:extraction} using tweet from Twitter as an example.

\subsection{Ingestion Service} \label{sub:apibase_ingestion}

We implemented a set of micro-services (Twitter) to obtain and persist data for further use within a data lake,
CoreDB~\cite{beheshti2017coredb}. This enables us to deal with dynamism of the data arrivals and also large sets of social media data. Then, we define a schema for information items and persist them MongoDB (a data island in our data lake) in JSON\footnote{https://json.org} format. JSON is a popular and simple to parse text format for data interchange.

\begin{figure}[t]
\centering
\includegraphics[width=4.5in,height=\textheight,keepaspectratio]{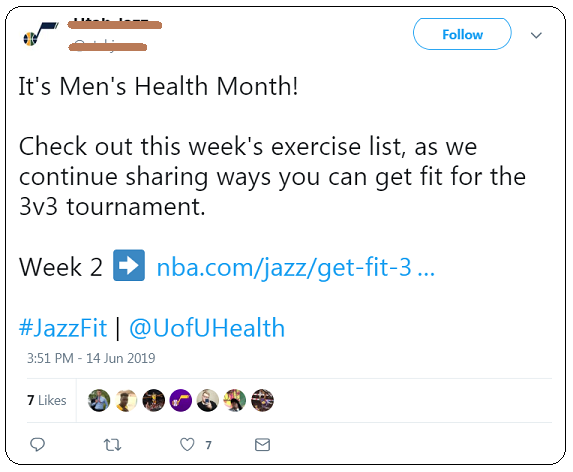}
\caption{A tweet ingested from Twitter.}\label{fig:ingestT}
\end{figure}

Each tweet within Twitter contains several attributes. As an example, refer to Figure~\ref{fig:ingestT} for a tweet. Some of the important attributes ingested from the tweet are discussed are:

\begin{compactenum}
\item{\textbf{Text}} - Text within a tweet;
\item{\textbf{Hashtags}} - List of hashtags within the tweet e.g., \#JazzFit;
\item{\textbf{Links}} - List of links mentioned within the tweet e.g., nba.com/jazz/get-fit-3;
\item{\textbf{User}} - Name and other details of the user e.g., UtahJazz;
\item{\textbf{Geo}} - Location from where the tweet was posted e.g., Utah, USA.
\end{compactenum}

Post the ingestion process, the raw tweets in JSON format are available for further use. An example tweet stored in JSON format is illustrated below in Figure \ref{fig:tweet_ingested}.

\begin{figure}[h!]
\centering
\includegraphics[width=\textwidth,height=\textheight,keepaspectratio]{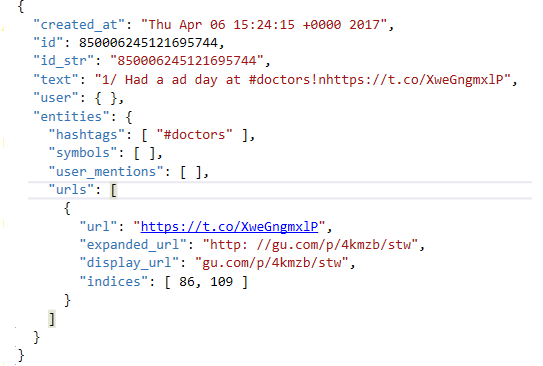}
\caption{Tweet stored in JSON format.}\label{fig:tweet_ingested}
\end{figure}

\subsection{Extraction Service}
\label{sub:impl_graphdb}

Next in our curation pipeline, we design and implement services to extract items from the raw data. These items consists of features which are of value for driving meaningful references. These features include:

\begin{compactenum}
\item{\textit{Lexical features}}: These are words that form part of vocabulary of language. This includes keywords, misspellings, abbreviations and slangs. For example from the tweet in Figure~\ref{fig:ingestT}; \textit{tournament} would be extracted as a keyword.
\item{\textit{Natural Language features}}: Words that can be extracted from analysis of natural language such as named-entities (e.g., person name, organisation, product etc.) and part-of-speech (e.g., noun and verb). For example from the tweet in Figure~\ref{fig:ingestT}; \textit{Men} would be extracted as a noun.
\item{\textit{Time and Location features}}: Mention of time and location in the social media post, i.e., a tweet. For example in Twitter, tweet may contain the location of posting. For example from the tweet in Figure~\ref{fig:ingestT}; ``\textit{14 June 19}'' would be extracted as date.
\end{compactenum}

To sum up, we perform data curation feature engineering by identifying variables that encode information for analytics. We extract these variables for cleansing and curation further down the pipeline. The extracted features are stored in a \textit{featureDB} using Microsoft SQL server database engine.

\section{Automated Curation : Extraction and Correction}
\label{sec:apibase_automated}

This section presents the architectural overview of automated correction step for CrowdCorrect pipeline.

In this step, we leverage external knowledge sources and services to automatically correct the data. It is important to note that, we perform curation and correction of extracted features; variables that encode information and help derive meaningful inferences. We term this as data curation feature engineering. Examples of features extracted from tweet text are keywords, named-entities and so on. This is further discussed in the following section.

\subsection{Automated Correction Services}
\label{sub:apibase_impl_knowledge_graph}

Once the extracted features are available; we implement services to automatically identify and correct those features. We focus on correcting misspellings, jargons (i.e., special words or expressions used in a professional context, which are difficult to understand) and abbreviations using external knowledge sources; available as services as illustrated below in Figure~\ref{fig:external-automated}.

\begin{figure}[h!]
\centering
\includegraphics[width=\textwidth,height=\textheight,keepaspectratio]{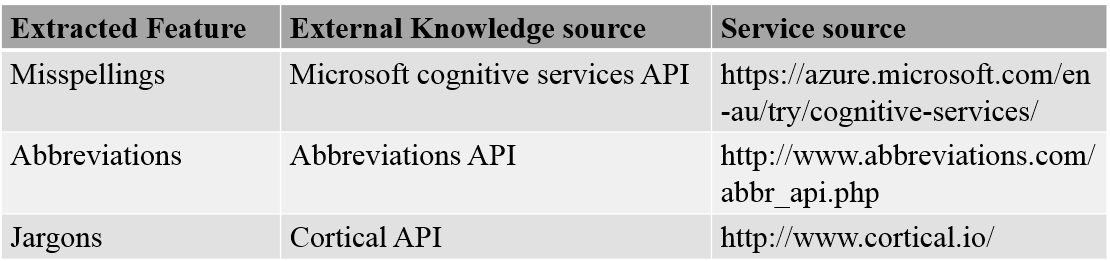}
\caption{List of external knowledge source for each feature.}\label{fig:external-automated}
\end{figure}

This automated correction step does the following. It submits each extracted feature to each of the three external services (shown in Figure~\ref{fig:external-automated}). The services return with a matching words with scores. For example, for the misspelled word ``\textit{healht}''; the Microsoft Cognitive service API\footnote{https://azure.microsoft.com/en-au/services/cognitive-services/spell-check/}, returns the word ``\textit{health}'' with a score of 1. Similarly, the abbreviations API\footnote{http://www.abbreviations.com/abbr/api.php} and Jargons API\footnote{http://www.cortical.io/} return likely matches with scores. The automated correction step outputs cleaned and corrected raw data in an annotated dataset format.

\section{Crowd Curation : Correction Tasks}
\label{sec:apibase_automated}

In this step, we design a simple web interface to facilitate users in the crowd to correct items which were not corrected in the last step. To achieve this goal, we design two micro-tasks namely suggestion and correction micro-tasks. Both these tasks are automatically generated and presented in a web-interface. To automatically generate a micro-task; we designed a heuristic, which determines what task to present to the user. Our goal is to have a hybrid combinations of crowd workers and automated techniques such that we can build collective intelligence. The design of the two types of micro-tasks are illustrated in the next sub-section.

\subsection{Crowd Tasks Generation}
\label{sub:apibase_impl_crowd_tasks}

The core of the crowd-sourced correction relies on crowd tasks generator service. This service accesses the annotated dataset and builds a simple micro-task in form of multiple choice question and answer format. The two key types of tasks generated include:

\begin{compactenum}
\item \textit{Suggestions} - A micro-task for the user to select if the presented feature within a tweet is a jargon, abbreviation or misspelling.
\item \textit{Correction} - A micro-task for the user to select the possible match for a presented tweet, keyword (feature) and issue (jargon, abbreviation or misspelling).
\end{compactenum}

Possible answers for the suggestions tasks are jargon, abbreviation, misspelling or none. For example, we present a tweet; ``Hosp. are running short on trained doctors''. Along with a tweet, we also present the crowd user with a question if the keyword ``Hosp.'' is a jargon, abbreviation or a misspelling. A user can select their answer by click of a radio button on the web page. Further, we also have a option of selecting none for cases where there are no issues.

Similarly possible answers for corrections are sourced from the external knowledge sources used earlier for misspellings, jargons and abbreviation. For example, from the earlier suggestion question ``Hosp. are running short on trained doctors''; we would have possibly verified ``Hosp.'' to be an abbreviation. Then in a correction question, we present the user with a question to provide us a full-form. We present a range of options sourced from external knowledge sources (abbreviations API, in this case). In addition, we also allow a user to type in, if they desire to do so in a free text field. The correct answer in this example would be ``hospital''. The interfaces for these tasks are illustrated further in the next chapter.

\subsubsection{Suggestion Micro-tasks}
\label{sub:apibase_impl_suggestions}

\begin{figure}[h!]
\centering
\includegraphics[width=\textwidth,height=\textheight,keepaspectratio]{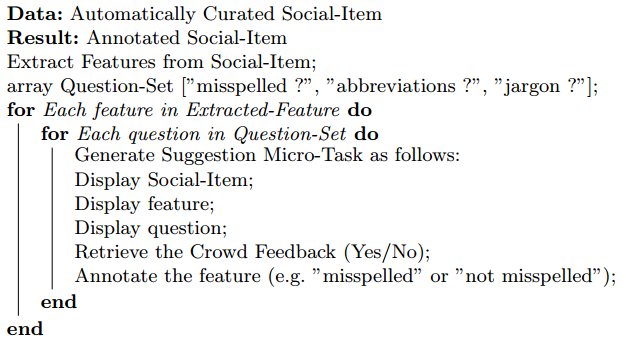}
\caption{Algorithm for automatically generating suggestions micro-tasks.}\label{fig:suggestions_algo}
\end{figure}

We design and implement an algorithm to present a tweet with an extracted feature(e.g., keyword) to ask the crowd user if the extracted feature can be considered as misspelled, jargon or abbreviation. An illustration of this is shown in Figure~\ref{fig:suggestions_algo}.

\subsubsection{Correction Micro-tasks}
\label{sub:apibase_impl_corrections}

\begin{figure}[h!]
\centering
\includegraphics[width=\textwidth,height=\textheight,keepaspectratio]{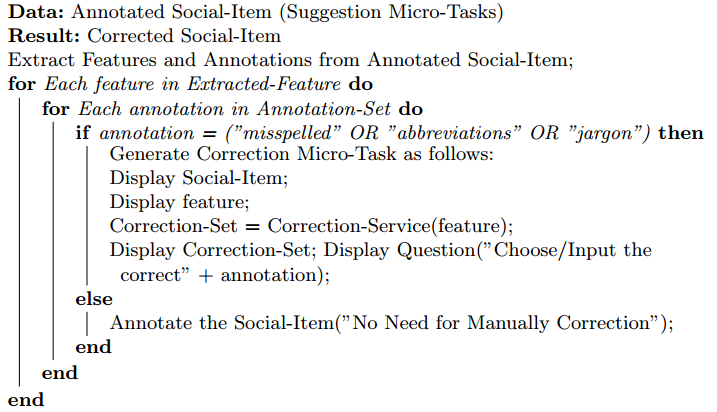}
\caption{Algorithm for automatically generating correction micro-tasks.}\label{fig:corrections_algo}
\end{figure}

We design and implement a corrections algorithm for users to select the correct form of a feature. For example, if a tweet's feature (keyword) is identified as a abbreviation; we automatically generate correction matches and present it to them to select the most appropriate. The automatic generation of correction matches relies on the the external knowledge sources (services) we mentioned earlier. An illustration of a correction micro-task is presented in Figure~\ref{fig:corrections_algo}.

\section{Conclusion}
 \label{sec:apibase_conclusion}

In this chapter, we discussed the challenges in analysing raw social data. Particularly we looked at issues that arise in social media data due to unusual syntax and style of text. We looked at normalization and contextualization techniques to improve the quality of text data. However, due to the range of problems encountered such as out-of vocabulary words, abbreviations and slangs. cleansing and curation of social media text remains a challenge.

We address the above challenge by proposing CrowdCorrect, an extensible curation pipeline. CrowdCorrect perfoms automated curation of features, followed by crowd-sourced approach to correct features which failed in the automated step.

\chapter{Implementation and Evaluation}\label{chp:imp_evl}

In this chapter, we discuss the implementation of the proposed cleansing and curation pipeline, CrowdCorrect. We summarise the motivational challenges behind the research and development. We detail microservices developed for ingestion, extraction and correction of the raw social data. We present a motivating scenario to evaluate the CrowdCorrect pipeline. Our chosen use case consists of tweets from Twitter in advent of Australian Government's budget announcement. Finally, we design and run an experiment using the raw social media data from the motivational scenario. To conclude, we present the evaluation results of the experiment.

\section{Introduction} \label{sec:imp_evl_introduction}

Social network sites by design empower their users to express and share their ideas, thoughts and opinions to a wider audience. This has lead to an exponential rise in popularity of the social media sites such as Twitter and Facebook \cite{van2013understanding}. The data within these social channels natively captures the beats of the masses \cite{stieglitz_social_2018}. This has opened up new opportunities for deeper understanding of several aspects such as trends, opinions and influential actors. This social media data provides valuable insights to aid decision making in diverse areas such as marketing, public policy and healthcare. Therefore, analytics of social media data is considered as vital and strategic priority for organisations and government.

Raw data from social platforms is generally semi-structured and noisy \cite{soto_data_nodate}\cite{eisenstein2013bad}. Such noise can include misspellings, slang words, abbreviations, truncations, incorrect syntax and grammatical errors. To sum up, the quality of the raw social data is low  \cite{immonen2015evaluating}; which introduce linguistic challenges in algorithms for analytics and can lead to inaccurate analysis \cite{adedoyin-olowe_survey_nodate}. Therefore, there is a need to transform raw data into contextualized data and knowledge. This transformation process is referred to as data curation and cleansing forms an integral part of it. Next, we look at our proposed cleansing and curation pipeline for social media data, namely CrowdCorrect.

CrowdCorrect is an extensible social media data cleansing and curation pipeline. The key focus for the pipeline is to cleanse raw social data; using both automated and crowd-sourced techniques. The pipeline consists of set of micro-services that also leverage external knowledge sources and services. An illustration of this pipeline was presented in  Chapter 3 earlier. The micro-services are broken down into three activities namely pre-processing, automatic correction and crowd sourced correction. The key motivation for the development of CrowdConnect, is the low quality raw data on social sites such as Twitter. The quality challenges posed by raw social data and the difficulty faced by automated techniques lead us to leverage crowd-sourcing approaches.

 In order to understand these challenges and evaluate CrowdCorrect, we present a motivating scenario in the next section. Further on, we discuss the implementation of CrowdCorrect and experimentation using the motivating use case.

\section{Motivating Scenario} \label{sec:imp_evl_motivatingusecase}
\begin{figure}[h!]
\centering
\includegraphics[width=\textwidth,height=\textheight,keepaspectratio]{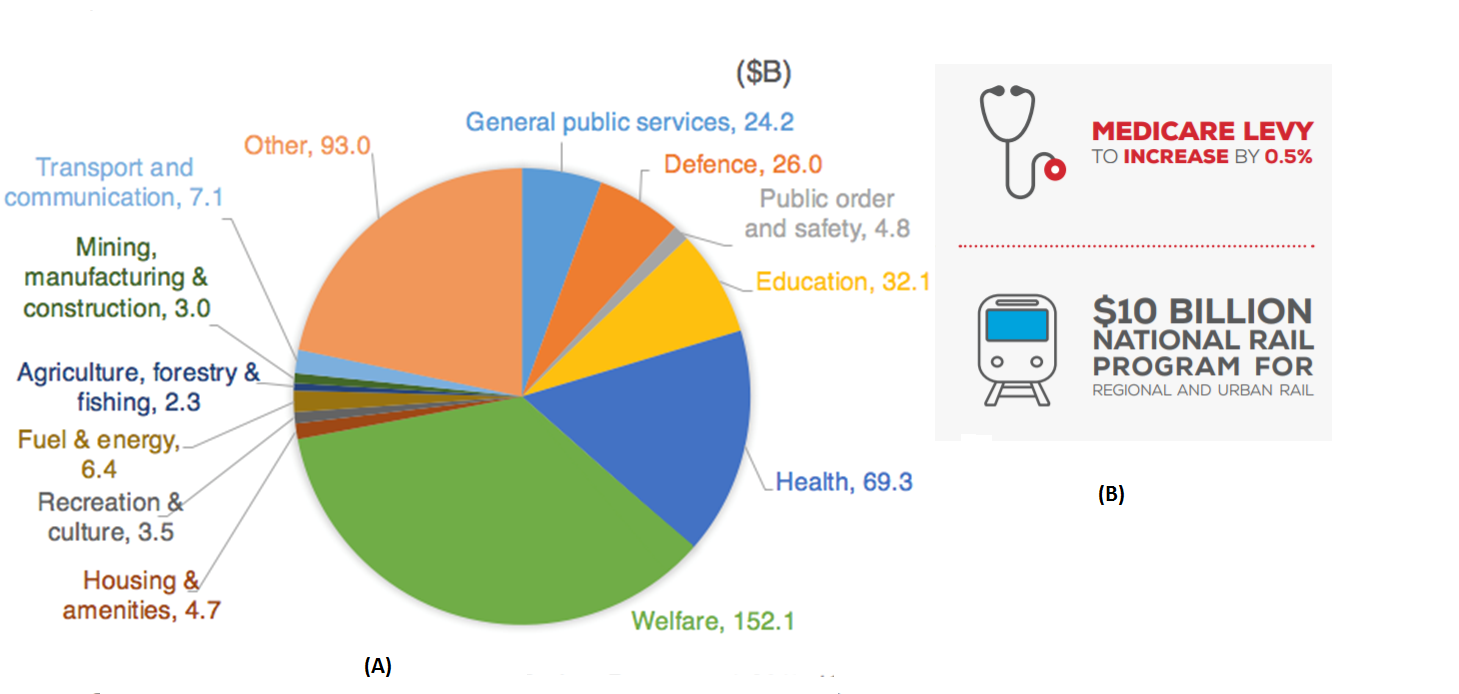}
\caption{An example of budget categories (A) and associated programs~ (B)\protect\footnotemark}\label{fig:budget}
\end{figure}

In order to evaluate our CrowdCorrect pipeline, we looked for potential use cases within the social media channels. Then, we narrowed down on a use case containing a corpus of tweets from Twitter in advent of budget announcement by the Australian Government. The key criteria for selection of a use case was presence of large number of issues that consisted of  usage of slang words (jargons and abbreviations) and misspellings. Further, we also considered an analytics task related to  ``understanding
Government's Budget in the context of Urban Social Issues''. A typical governments' budget denote how policy objectives are reconciled and implemented in
various categories and programs. Figure \ref{fig:budget}(A) shows the overall budget categories for the 2017 Federal budget.

\footnotetext{https://www.charteredaccountantsanz.com/news-and-analysis/news/2017-18-australia-budget}

Budget categories (e.g., Health, Social-Services,  Transport and Employment) are then broken down into hierarchical set of programs (e.g., Medicare Benefits in Health, and Aged Care in Social-Services).
These programs refers to a set of activities or services that meet specific policy
objectives of the government \cite{kim2016budgetmap}. An example of such programs are shown in Figure \ref{fig:budget}(B). Social media channels are abuzz with reactions pertaining to government's budget announcements. In order to accurately guage public opinion on various programs related to budget; an analyst would tend to first classify social media feed into the various categories respectively. This would be difficult using traditionally adopted budget systems,
which may make it difficult to accurately evaluate the governments' services requirements and performance.

\section{Methodology} \label{sec:imp_evl_method}

As the first step, we analyze the different budget categories from our selected use case. Since there are many categories (e.g., health, defence and social welfare), therefore for the purpose of cleansing and curation, we picked the ``\textit{health}'' category. That is, identifying and curating tweets related to \textit{health} category within a corpus of budget related tweets. The key reason for picking health was also tied to the various key government spending initiatives with regards to medicare\footnote{https://www.humanservices.gov.au/individuals/medicare
} and hospital treatments in the budget announcement. These health related initiatives are always a constant source of debates in the popular media, leading to increased public attention and scrutiny \cite{smith2017positioning}\cite{sowa2018private}.

There are number of issues in social media data, some of which we highlighted in Figure \ref{social-text}. We selected three classes of textual issues which we could use as a basis for cleansing and curating tweets. Namely, they are jargons, abbreviations and misspellings. Jargons and abbreviations are part of slang words used widely on the Internet. Together, these three classes are the most popular types of issues found in social media channels such as Twitter \cite{zappavigna2012discourse}\cite{mosley2012social}\cite{liu2012broad}. The key challenge for an analyst in such a scenario would be to be able to cleanse and classify tweets correctly in the health category and associated government programs.

Then, as discussed in the following sections, we designed and implemented a set of micro-services and a user interface, which would collectively form parts of the CrowdCorrect pipeline (Figure \ref{fig:crowd-correct}). This is discussed in  section \ref{sec:apibase_implementation} in more detail. The micro-services are for ingestion, automated and crowdsourced cleansing and curation.

Finally, to evaluate our pipeline, we designed an experiment to automatically correct tweets and then engage crowd users to help us further cleanse and curate the tweets, as discussed in the Section \ref{sec:evaluation}.

\section{Implementation}
\label{sec:apibase_implementation}

Implementation consists of building a pipeline of activities starting with:\begin{inparaenum}[(i)] \item ingestion of raw tweets; \item followed by extraction of keywords (features); \item micro-services to automatically cleanse the tweets; \item building a crowd facing front-end tool and related services; and \item Prepare the curated data set\end{inparaenum}. After these steps, machine learning algorithms may be used to classify the tweet as related to specific budget category. As we discuss the details in subsequent sections, the pipeline persists raw ingested social media data using MongoDB\footnote{https://www.mongodb.com/
} for raw data and SQL Server\footnote{https://www.microsoft.com/en-au/sql-server} for curated result set. Microservices for automatic and crowd cleansing were developed using the Microsoft's .NET framework\footnote{https://dotnet.microsoft.com/download/dotnet-framework/net472}. In addition, we leverage existing services \cite{beheshti2017automating} for the purposes of data ingestion and extraction of features from raw data. We discuss each step in detail, in the following sections.

\subsection{Data Ingestion and Extraction of Features}

At this initial step, we import social data using micro-services and store them inside MongoDB in JSON\footnote{https://www.json.org} format. In the budget scenario, the Treasurer announced the budget on Tuesday 3 May, 2016. We collected all tweets from one month before and two months after the budget announcement. This comprised of about 15 million raw tweets which were persisted and indexed inside our MongoDB data store. They key fields within the persisted tweets for us are text and hashtags (Section \ref{sub:apibase_ingestion}), as we build upon them to perform the cleansing task.

Following on, we perform feature extraction using existing open source services \footnote{https://github.com/unsw-cse-soc/Data-curation-API
} \cite{beheshti2017automating}. In particular, we leverage the \textit{keyword} extraction service. Keywords are words for great importance or value. From a scientific perspective, keywords help to filter and index data. In essence, we cleanse and curate keywords within text; as they are candidates for machine-learning classifiers to classify items into appropriate classes. Table \ref{keyword} shows an example of a tweet (with misspelling and abbreviation) and extracted keywords using our services. At the end of this step, we would have extracted all the keywords from raw tweets.

\begin{table}
\centering
\begin{tabular}{|p{6cm}|p{3cm}|}
\hline
 \textbf{Tweet} & \textbf{Extracted Keywords}\\ \hline
 My cardio won't like the govt plan on hulthcare \#ausbudget & cardio, govt, plan, ausbudget, hulthcare  \\
 \hline
\end{tabular}
\caption{Keywords extracted from tweet.}
\label{keyword}
\end{table}

\subsection{Automated Correction Microservices} \label{sec:imp_evl_microservices}

Now, we develop a set of micro-services to automatically correct keywords from the tweets. In order to achieve this goal, we link extracted information to external knowledge base and services as shown in the Table \ref{services-used}.

For misspellings, cleansing services replace keywords with the possible match with highest score. For example, the Bing Spell check returns a correct spelling with a statistical score. Similarly for abbreviations, we perform the same process. For jargons, we developed a list of standard forms of words related to health category of the budget. This is form of background knowledge or meta-data for a given use case. For example, both \textit{cardiologist} and \textit{neurologist} may refer to the term \textit{doctor}. Our services inspects jargons and then matches them to the standard forms and if a match is found, then a replacement takes place.

\begin{table}
\centering
\begin{tabular}{|p{4cm}|p{3cm}|p{7cm}|}
\hline
 \textbf{Service} & \textbf{Purpose} & \textbf{Link}\\ \hline
Microsoft Bing Spell check & Identify misspellings & $https://azure.microsoft.com/en-au/services/cognitive-services/spell-check/$\\
\hline
Abreviations API & Identify a word as an abbreviation and get the full form & $http://www.abbreviations.com/abbr$\\
\hline
Jargons & Find matching words & $http://www.cortical.io/$\\
 \hline
\end{tabular}
\caption{Reference for External Services.}
\label{services-used}
\end{table}

It is important to note here, that since the automated services rely on best scores from external services to replace keywords, this is likely to introduce errors or wrong word matches. For example, we checked score for a misspelled word \textit{cardo} against Bing Spell check from a tweet. Bing's best match was \textit{card} with score of about 90\%. Ideally in this scenario, the correct word would be \textit{cardio}. Therefore, crowdsourcing can help us identify and correct issues, which failed to be rectified in the automated step.

\subsection{Crowdsourced Correction} \label{sec:imp_evl_microtasks}

\begin{figure}[h!]
\centering
\includegraphics[width=\textwidth,height=\textheight,keepaspectratio]{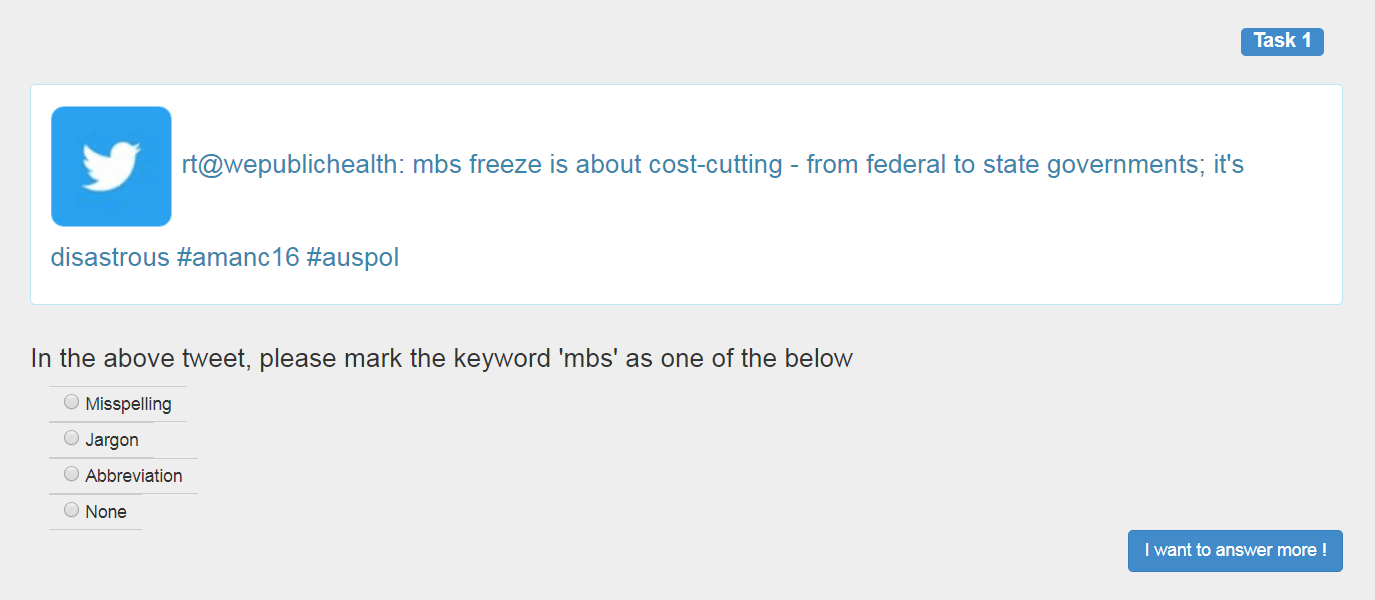}
\caption{An example of a Suggestion Crowd Task generated from our tool.}\label{fig:crowdtasks}
\end{figure}

\begin{figure}[h!]
\centering
\includegraphics[width=\textwidth,height=\textheight,keepaspectratio]{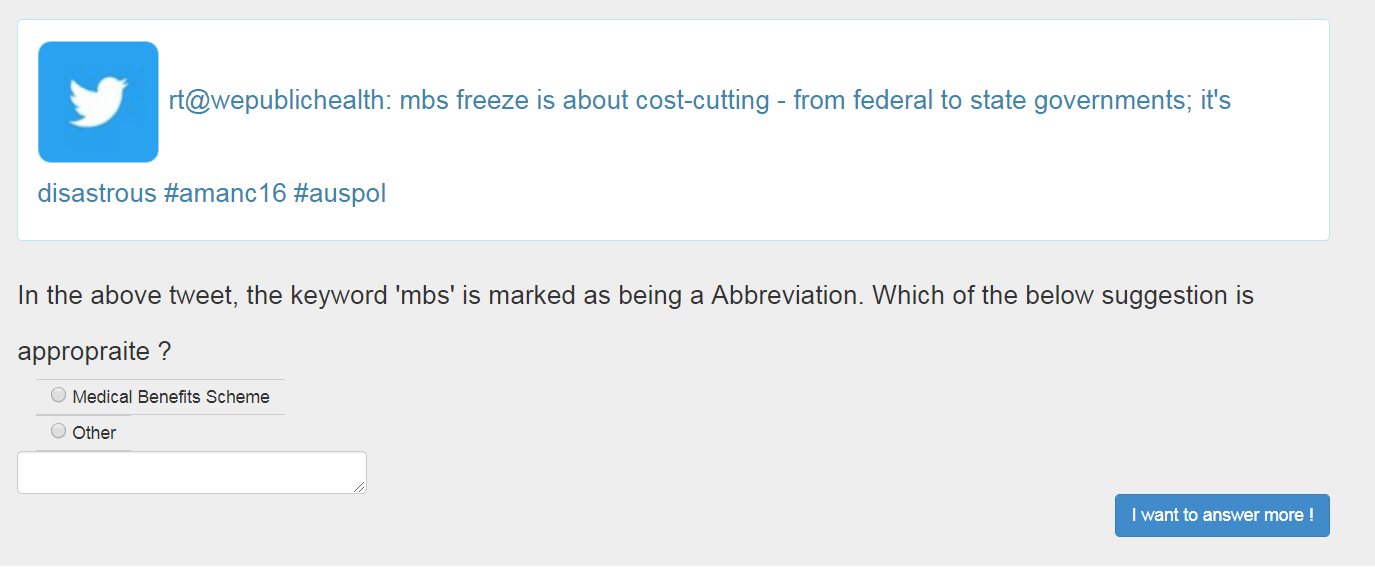}
\caption{An example of a Correction Crowd Task generated from our tool.}\label{fig:crowdtasks1}
\end{figure}

In this step, we developed a simple web-based user interface along with a set of micro-services. This we interface can be accessed via a web browser such as Chrome\footnote{https://www.google.com/chrome/}. Each crowd task consists of ten questions with multiple choice of answers, which a user can answer with a simple click. An illustration of crowd tasks generated from our pipeline is shown in Figures \ref{fig:crowdtasks} and \ref{fig:crowdtasks1}.

There are three types of questions we pose to a crowd user:

\begin{compactenum}
\item \textbf{Identification}: Identify if a tweet is related to the health category or not ?
\item \textbf{Suggestion}: Suggest if a keyword is a misspelling, jargon or an abbreviation. As an option, a user can choose none if there are no matches.
\item \textbf{Correction}: Select the best answer for correction or option to write your own.
\end{compactenum}

The identification task helps us narrow down the tweets which belong to health category. We have used this to filter out tweets from the initial 15 million tweet dataset. The suggestion tasks help us to identify if a particular keyword is a misspelling, jargon or abbreviation. Therefore, there are one of these three options for a crowd user to choose from. In order to present a suggestion task, we developed a heuristic which ensures we get maximum tweet coverage. In the context of the heuristic, a \textit{social-item} is a piece of data such as a tweet.

Once, we identify a particular keyword within a tweet, as a misspelling for example, we once again leverage the same set of services (refer to Figure ~ \ref{services-used}) to present options to the user. We use a simple algorithm of high score from all crowd user's answers to judge if a particular keyword is a jargon, keyword or abbreviation. The psuedocode for the correction heuristic, shown in Figure \ref{fig:suggestions_algo} in the chapter 3.

The result of the crowd cleansed and curated tweets were persisted inside Microsoft SQL Server database\footnote{https://www.microsoft.com/en-au/sql-server/sql-server-2017
}. We ran a simple heuristic of selecting the best result based on majority vote score. An illustration of this heuristic is highlighted in  \ref{fig:corrections_algo} in Chapter 3.

\section{Evaluation}
\label{sec:evaluation}

We have used three months of Twitter data from May 2016 to August 2016 which was roughly fifteen million tweets. The size of raw tweets was large and also had large number of tweets not related to the health catefory, therefore, we first ran a crowdsourced identification task, after ingesting the tweet data inside MongoDB. The aim of this task was for a crowd user to simply identify if a tweet belonged to health category or not.

Following on, we run an experiment  using identified tweets from our use case from the earlier step. This experiment included extracting keywords, automated correction and finally crowdsourced correction. In order to demonstrate the effectiveness of the CrowdCorrect approach, we created two datasets in the field of healthcare. Raw tweets forms the first part of dataset; while curated tweets; where all jargons, misspellings and abbreviations were corrected, formed the other datset.

The following steps summarise our methodology used to perform an experiment: \begin{inparaenum}[(i)]
\item Data Ingestion - ingestion three months of Twitter data which included time before and after the budget announcement, \item Identification crowd task - where we asked crowd users to identify if a tweet belongs to health or not, \item Keyword extraction - extraction of keywords from the tweet, \item Automated correction and finally and \item Crowdsourced correction - based on generating crowd micro-tasks.\end{inparaenum}

Finally for evaluation of our approach, we developed four machine learning classifier using a binomial logistic regression and a gradient descend algorithm. A logistic regression classifier is a generalized linear model that we can use to
model or predict categorical outcome variables. On the other
side, gradient descend algorithm is widely used in optimization problems, and
aims to minimize a cost function. The classifiers were trained to match tweets against two classes namely: \textit{health} or \textit{Other}. That is, we wanted to evaluate the effectiveness of cleansing and curation operations of the proposed approach.

\subsection{User Selection for Crowdsourced Tasks}
\label{sub:crowd_selection}

In our evaluation of the experiment, we asked students enrolled in semester two, 2017 in the Web Application Engineering unit\footnote{www.cse.unsw.edu.au/~cs9321/16s2} to be the main participants as the crowd users. In addition, we also encouraged members of the service oriented computing group at the University of New South Wales to be members of the crowd. Finally, we also invited a set of crowd users from a local organisation\footnote{http://www.westpac.com.au} to be part of the experiment in 2018. The total strength of our crowd users was close to 500 people.

Each potential participant gets an invitation via an email. A web link in the email navigated the user to the web interface for crowd micro tasks, after providing two unique identifiers: \textit{name} and \textit{email address}. Each invitation email also had information which were number of questions to answer, duration and a help guide.

\begin{figure}[h!]
\centering
\includegraphics[width=\textwidth,height=\textheight,keepaspectratio]{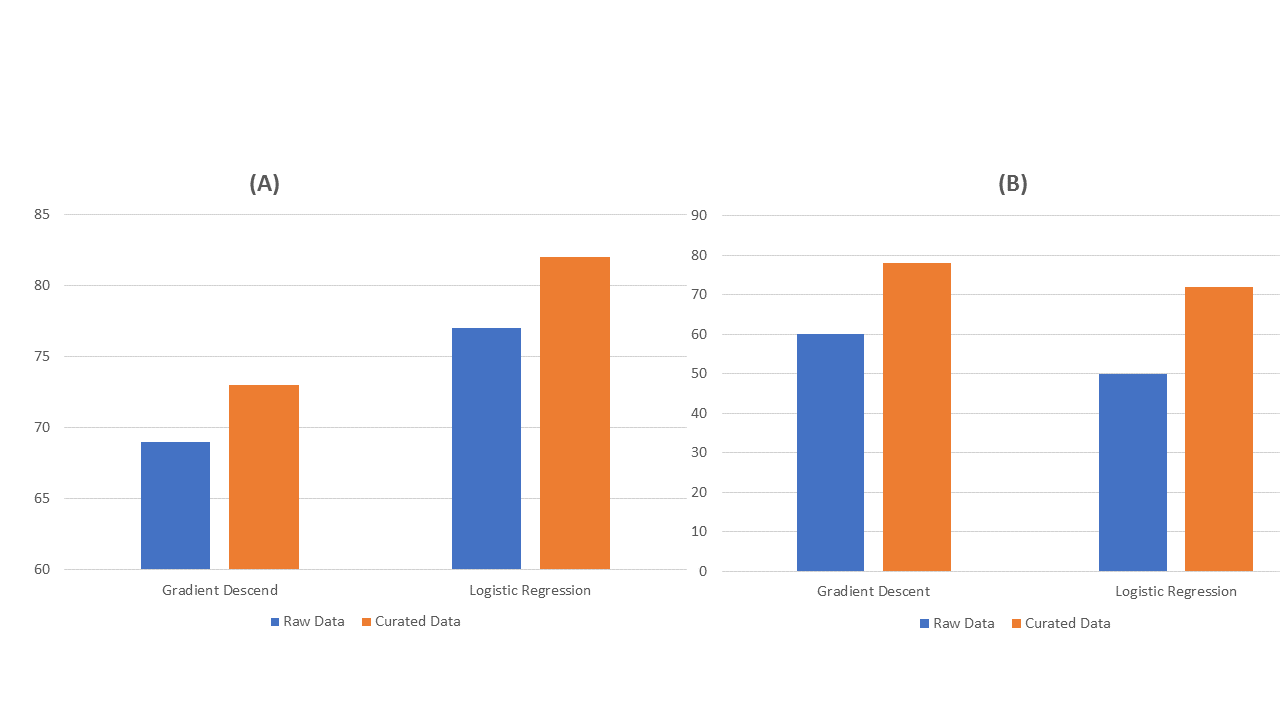}
\caption{Results of experiment run - comparison of Raw and Curated data.}\label{fig:results}
\end{figure}

\textbf{Discussion.}
 In this experiment, the classifiers have been constructed to verify if
a tweet is relevant to health category or not. First we trained two classifiers (Logistic
Regression and gradient descend algorithms) using the raw and curated tweets.
For training classifiers, we filtered out tokens occurred for less than three times.
We also removed punctuation and stop words. We used porter stemmer for
stemming the remaining tokens. The results of our experiment are summarised in the Figure \ref{fig:results}. Both logistic regression and gradient descend algorithm outperformed in the curated dataset.
In particular, the gradient descend algorithm has improved the precision by 4\%,
and the amount of improvement using the logistic regression algorithm is 5\%. In
addition, Figure 4(B) illustrates the measure improvement in F-measure: the Fmeasure has improved in both gradient descend classifier and logistic regression
classifier by 2\% and 3\% respectively.

\section{Conclusion}

In this chapter, we detailed our implementation of CrowdCorrect cleansing and curation pipeline. We discussed on how CrowdCorrect ingests data and then extracts keywords from tweets to perform further cleansing on them. Further, we illustrated how the automated and crowdsourced activities leverage the external knowledge bases and services. We also illustrated how we have developed tools to engage the crowd and gather feedback and use it for correction of the raw data. The end result or output from utilizing our approach is a curated set of data which can then be fed for further analytics. Although, we illustrated one use case with Australian Government Budget tweets; this approach can be used for other use cases as well. Our experimental results illustrate the effectiveness of using curated dataset over raw data for any reliable analytics task.

Our approach leverages the collective wisdom of the crowd to improve the quality of raw data leading to a more robust curation. Essentially, it adds another layer of quality check and correction where automated approaches struggle. One downside of our approach is the engagement of the crowd ,i.e., selection, motivation and unbiased participation of the crowd users which needs to be addressed.
\chapter{Conclusion and Future Work}
\label{chp:conclusion}

In this chapter, we summarise the contributions of our research work and discuss future opportunities to extend further on the research work.

\section{Concluding Remarks}

Social media sites have grown rapidly since their first introduction in 2000s\cite{chaffey2016global}. The popular sites such as Twitter\footnote{http://twitter.com}, Facebook\footnote{http://facebook.com} and Instagram\footnote{https://www.instagram.com}, have combined user populations that run into billions\footnote{https://www.smartinsights.com/social-media-marketing/social-media-strategy/new-global-social-media-research/}. Further, governments and organisations have also taken to the social media examine their policies \cite{bertot2012impact}\cite{hagen2018government}, develop products and guage sentiments \cite{jeong2017social}, marketing strategies  \cite{tuten2017social} and so on. 

Naturally, due to the immense popularity of social media channels, a lot of valuable data is published on a daily basis. This social data is openly available to be queried; and analytics of such social data has become a vital priority for organisations and government. However, there are many roadblocks to utilize this data for valuable purpose. As such, this data is constantly flowing (velocity) and is large (volume). In our research work, we examined and proposed a framework for another major roadblock, which is the quality of this data. 

In this thesis, we discussed that due to non-standardization of use of language on social sites such as Twitter, making sense of the data is difficult. Raw social data usually contains misspellings, slangs and lack the use of proper grammar\cite{soto_data_nodate}\cite{eisenstein2013bad}. Data Cleansing and Curation offers a potential solution to cleanse raw social data. Automated cleansing techniques perform poorly against social media data\cite{adedoyin-olowe_survey_nodate}. To address these challenges we proposed an extensible curation and cleansing framework, CrowdCorrect. In our proposed curation framework, we embedded a crowdsourced approach or crowd cleansing in addition to automated techniques. We discussed the motivations and rationale of building a cleansing and curation pipeline CrowdCorrect in Chapter 1. Below we summarise the contributions of our research:

\begin{compactenum}[]
	\item \textbf{Study of State of the Art}. At first, we looked at research work for social media analytics. We found that a lot of research was focussed on specific attributes (features such as Like in Facebook) or research on social media sites themselves. There isn't a significant body of research looking at solving quality issues in social media data for analytics. Further, we discussed what data cleansing and curation are and also discussed the various curation techniques and frameworks in literature. We found that a lot of existing frameworks either cater for structured data or do not contain an end-to-end pipeline with a focus on cleansing. Then, we discussed research work and applicability of crowdsourcing techniques, also specifically with social media data.
	\item \textbf{CrowdCorrect}. We proposed an extensible cleansing and curation pipeline for social media data. This pipeline ingests, extracts raw data and features from social media sites. Further, we discussed the use of automated and crowdsourcing techniques to cleanse and curate the raw social data. In order to achieve this, we leveraged external knowledge sources and services. CrowdCorrect pipeline has two major activities:
	
	\begin{inparaenum}[(i)]

	\item \textbf{Automated feature extraction and correction} - We discussed the design and implemention of micro-services to extract features such as keywords from a corpus of tweet data and automatically perform major data cleansing tasks on extracted keywords. \\
		\item \textbf{Crowdsourced correction} - We discussed our approach to then use crowd inputs to further cleanse data which could not be corrected in the earlier step. In order to achieve this, we take extracted features (e.g., keywords) from the earlier step and automatically generate micro-tasks with possible options for the user to choose from. These micro-tasks are presented to users within a simple web interface. Our micro-tasks generation service uses external knowledge bases such as Bing spell check\footnote{https://azure.microsoft.com/en-au/services/cognitive-services/spell-check/} to suggest possible answers. 
		\end{inparaenum}
\end{compactenum}

\section{Future Directions}

Given the vitality of analytics of social media data, there is a lot of possible future work to extend the research. In our research work, we focussed on extracted keywords for cleansing and curation. As a future work, this can be extended to look at other extractable features within the raw social media text such as named entities, topics and sentiments. The pipeline can also be extended to automatically use tools such as Twitter Polls\footnote{https://help.twitter.com/en/using-twitter/twitter-polls
} to passively engage crowd users.
In addition, as an ongoing and future
work, we propose designing micro-tasks to turn the knowledge of the domain expert
into a domain mediated model presented as a set of rule-sets to support cases
where the automatic curation algorithms and the knowledge of the crowd may
not able to properly contextualize the social items.

\bibliographystyle{amsplain}
\bibliography{ms}

\end{document}